# A count probability cook book: spurious effects and the scaling model

S. Colombi[1,2], F.R. Bouchet[2], and R. Schaeffer[3]

[1] NASA/Fermilab Astrophysics Center, Fermi National Accelerator Laboratory,
P.O. Box 500, Batavia, IL 60510

[2] Institut d'Astrophysique de Paris, 98 bis boulevard Arago, F-75014 - Paris, France.

[3] Service de Physique Théorique, Laboratoire de l'Institut de Recherche Fondamentale du Commissariat à l'Energie Atomique, CEN-Saclay, 91191 Gif-sur-Yvette Cedex, France.

**Abstract**

We study the errors brought by finite volume effects and dilution effects on the practical determination of the count probability distribution function $P_N(n,\ell)$, which is the probability of having $N$ objects in a cell of volume $\ell^3$ for a set of average number density $n$. Dilution effects are particularly relevant to the so-called sparse sampling strategy. This work is mainly done in the framework of the scaling model (Balian & Schaeffer 1989), which assumes that the $Q$-body correlation functions obey the *scaling relation* $\xi_Q(\lambda \mathbf{r}_1, ..., \lambda \mathbf{r}_Q) = \lambda^{-(Q-1)\gamma} \xi_N(\mathbf{r}_1, ..., \mathbf{r}_Q)$. We use three synthetic samples as references to perform our analysis: a fractal generated by a Rayleigh-Lévy random walk with $\sim 3.10^4$ objects, a sample dominated by a spherical power-law cluster with $\sim 3.10^4$ objects and a cold dark matter (CDM) universe involving $\sim 3.10^5$ matter particles.

The void probability, $P_0$, is seen to be quite weakly sensitive to finite sample effects, if $P_0 V \ell^{-3} \gtrsim 1$, where $V$ is the volume of the sample (but $P_0$ is not immune to spurious grid effects in the case of numerical simulations from such quiet initial conditions). If this condition is met, the scaling model can be tested with a high degree of accuracy. Still, the most interesting regime, when the scaling predictions are quite unambiguous, is reached only when $n\ell_0^3 \gtrsim 30-50$, where $\ell_0$ is the (pseudo-)correlation length at which the averaged two-body correlation function over a cell is unity. For the galaxy distribution, this corresponds to $n \gtrsim 0.02 - 0.03 h^3$ Mpc$^{-3}$.

The count probability distribution for $N \neq 0$ is quite sensitive to discreteness effects. Furthermore, the measured large $N$ tail appears increasingly irregular with $N$, till a sharp cutoff is reached. These wiggles and the cutoff are finite volume effects. It is still possible to use the measurements to test the scaling model properties with a good accuracy, but the sample has to be as dense and large as possible. Indeed the condition $n\ell_0^3 \gtrsim 80 - 120$ is required, or equivalently $n \gtrsim 0.04 - 0.06 h^3$ Mpc$^{-3}$. The number densities of the current three dimensional galaxy catalogues are thus not large enough to test fairly the predictions of the scaling model. Of course, these results strongly argue against sparse sampling strategies.

**subject headings**: galaxies: clustering – methods: numerical – methods: statistical

# 1 Introduction

The observed galaxy distribution is generally admitted to be homogeneous at scales larger than $\sim 100h^{-1}$ Mpc (with $H_0 = 100h$ km s$^{-1}$ Mpc$^{-1}$). At small scales, however, it is strongly clustered. The measured two-body correlation $\xi_2(r)$ indeed exhibits a power-law behavior,

$$\xi_2(r) \simeq \left(\frac{r}{r_0}\right)^{-\gamma}, \quad r_0 \simeq 5h^{-1}\text{Mpc}, \quad \gamma \simeq 1.8 \tag{1}$$

(Totsuji & Kihara 1969, Peebles 1974), on a large scale range ($0.1 \lesssim r \lesssim 10h^{-1}$ Mpc). The measurement of higher order correlation functions $\xi_Q(\mathbf{r}_1, ..., \mathbf{r}_Q)$ seems moreover to indicate that they can be hierarchically decomposed as sums of products of $Q - 1$ terms in $\xi_2$ (Groth & Peebles 1977, Fry & Peebles 1978, Davis & Peebles 1983, Sharp et al. 1984) at least up to $Q = 8$ (Szapudi et al. 1992). This hierarchical model is a particular case of the more general *scaling relation* (Balian & Schaeffer 1988, 1989a, hereafter BS)

$$\xi_Q(\lambda\mathbf{r}_1, ..., \lambda\mathbf{r}_Q) = \lambda^{-(Q-1)\gamma}\xi_N(\mathbf{r}_1, ..., \mathbf{r}_Q). \tag{2}$$

It is commonly believed that the main source of galaxy clustering is gravitational instability. Both theoretical (Davis & Peebles 1977, Peebles 1980 hereafter LSS, Fry 1984a, Hamilton 1988, Balian & Schaeffer 1988) and numerical arguments (Efstathiou et al. 1988, Bouchet et al. 1991, hereafter BSD, Bouchet & Hernquist 1992, hereafter BH) indeed suggest that a system with gaussian initial fluctuations reaches a scale invariant behavior in the non-linear regime ($\xi_2 \gg 1$). The resulting hierarchy of correlations naturally depends on the initial conditions power spectrum. In the weakly non-linear regime ($\xi_2 \ll 1$), perturbation theory shows that a similar hierarchy appears with $\xi_Q \propto \xi_2^{Q-1}$ (LSS, Fry 1984b, Goroff et al. 1986, Grinstein & Wise 1987, Bernardeau 1991, Juszkiewicz et al. 1992), but the correlations hierarchy (i.e. the constants of proportionality) may be of course different from the one obtained in the highly non-linear regime (see, e.g., Colombi et al. 1994, hereafter CBS, Lucchin et al. 1994). A change in behavior is therefore expected at scales close to the correlation length, which does not seem to be the case at low $Q$ in the three dimensional observed galaxy catalogs (Bouchet et al. 1991, 1993, Gaztañaga 1992). Lahav et al. (1993, but see also Matsubara & Suto 1994) have recently argued that this could be due to projection in redshift space effects. Hence, a detailed study of the scaling model is an important stage of the understanding of large scale structure dynamics and statistics.

It is tempting to assume that the property (2) holds for any $Q$ and to see what are the consequences for the galaxy distribution. However, the direct measurement of the $Q$-body correlation functions becomes practically difficult and quite noisy when $Q \gtrsim 5$. Since the relation (2) reflects an homothetic property, it suggests to only look at the scaling behavior of the underlying distribution and forget the angular dependence of the correlation functions. The corresponding statistical tool is precisely the count probability distribution function $P_N(n, \ell)$ (hereafter CPDF). It measures, in a discrete set of average number density $n$, the probability that a cell of volume $v = \ell^3$ (or of size $\ell$), randomly thrown in the set, contains $N$ objects.

The CPDF is indirectly related to the averaged correlations $\overline{\xi}_Q(\ell)$, defined by

$$\overline{\xi}_Q(\ell) \equiv \ell^{-3Q} \int_v d^3r_1 ... d^3r_Q \xi_Q(\mathbf{r}_1, ..., \mathbf{r}_Q). \tag{3}$$

Indeed, its generating function $\mathcal{P}(\lambda) \equiv \sum \lambda^N P_N$ can be written (see, e.g., BS)

$$\mathcal{P}(\lambda) = \exp\left\{\sum_{Q=1}^{\infty} \frac{(n\ell^3)^Q(\lambda-1)^Q}{Q!}\overline{\xi}_Q(\ell)\right\}. \qquad (4)$$

The CPDF is simple to measure. It has several applications. The evaluation of the moment of order two of the CPDF, related to the averaged two-body correlation, indirectly estimates the effective clustering of the system as a function of scale and can be used to check the large scale homogeneity of the galaxy distribution. The moment of order three, related to the skewness of the distribution, is used to test gravitational instability in the weakly non linear regime (Juszkiewicz & Bouchet 1991, Juszkiewicz et al. 1993). The moment of order $q$ ($q$ real) of the CPDF provides indications on the multifractal behavior of the system (Balian & Schaeffer 1989b, Colombi et al. 1992).

Several models have been proposed to predict the shape of the function $P_N(n,\ell)$ and we quote here the most important ones. The oldest one is certainly the log-normal distribution (studied in detail by Coles & Jones 1991), since Hubble had already noticed in 1934 that it provided a good fit of the counts in cells measured on the projected galaxy distribution. Saslaw & Hamilton (1984), using a thermodynamic approach based on an assumption of statistical equilibrium, reached an analytical model where the CPDF depended on only one parameter $b(\ell)$. It was also seen to give a good description of the measured CPDF in the galaxy distribution (Crane & Saslaw 1986, 1988). Fry (1986) used the negative binomial model to fit the void probability of the galaxies distribution. This model was proposed by Carruther & Shih (1983) to explain particles multiplicities in high energy collisions. More recently, Balian & Schaeffer (BS), computed analytical predictions on the CPDF, assuming that the scaling relation (2) holds for any $Q$. They found that the function $P_N(n,\ell)$ should exhibit non trivial invariance properties that we now recall.

Consequences of the scaling relation have first been studied on the void probability $P_0(n,\ell)$ (hereafter VPDF) by White (1979) and Schaeffer (1984). They showed that in case the scaling relation applies, the function

$$\hat{\sigma}(n,\ell) \equiv -\frac{\ln(P_0)}{n\ell^3} \qquad (5)$$

should then depend only on the characteristic number $N_c$ of objects inside a cell located in an overdense region:

$$\hat{\sigma}(n,\ell) = \sigma(N_c), \qquad (6)$$

with

$$N_c \equiv n\ell^3 \overline{\xi}_2. \qquad (7)$$

This invariance property was successfully tested on the observed galaxy distribution (Sharp 1981, Bouchet & Lachièze-Rey 1986, Maurogordato & Lachièze-Rey 1987, Fry et al. 1989, Maurogordato et al. 1992), which suggests that equation (2) can indeed be generalized for all $Q$. Evaluating the VPDF in three numerical simulations, with cold dark matter (CDM), hot dark matter and white noise initial conditions, BH measured a slight disagreement with the scaling relation, but concluded that it could be attributed to misleading effects. Indeed, the apparent deviation was too small in comparison with the possible systematic and other ill-constrained errors. Vogeley et al. (1992) measured the VPDF on the extended CfA2 catalog, and claimed they find a significant deviation from the scaling relation at large scales. However, they did not properly take into account systematic errors due to finite sample effects, as it will be discussed in § 3. Before far-reaching conclusions are accepted, one indeed has to evaluate all the errors due to the unavoidable limitations of any realistic sample. Conversely, one should ask the following simple and practical questions: does a sample that should not verify the scaling relation always significantly disagree with equation (6)? Does a physical realization of an ideal scale invariant distribution always exhibit this property?

Similarly, as for the VPDF, if was found by BS that if the scaling relation applies, the function $P_N(n, \ell)$ scales in the strongly non-linear regime ($\overline{\xi}_2 \gg 1$) as an universal function $h(x)$ that only depends on one variable $x \equiv N/N_c$ instead of $N$, $n$ and $\ell$. More specifically, let us define the function $\hat{h}(N, n, \ell)$ by

$$\hat{h}(N, n, \ell) \equiv \frac{N_c^2}{n\ell^3} P_N(n, \ell). \tag{8}$$

Then, if equation (2) is verified and if $\overline{\xi}_2 \gg 1$, the function $\hat{h}$ can be written, in a certain regime,

$$\hat{h}(N, n, \ell) \simeq h(x), \quad x \equiv N/N_c. \tag{9}$$

The distribution function $h(x)$ is not arbitrary. It should present some asymptotic behaviors for $x \ll 1$ and $x \gg 1$, that we will detail in § 4. Its moments of order $Q$ are proportional to the ratios

$$S_Q(\ell) \equiv \frac{\overline{\xi}_Q}{\overline{\xi}_2^{Q-1}}, \tag{10}$$

that are constants with respect to scale if the scaling relation applies. Furthermore, the function $h$ is related to $\sigma(y)$ by the following transform

$$\sigma(y) = \frac{1}{y} \int_0^\infty (1 - e^{-yx}) h(x) dx, \tag{11}$$

so the functions $\sigma(y)$ and $h(x)$ theoretically involve the same information (but, practically, their determination is complementary, see, e.g., Bouchet et al. 1991, hereafter BSD, BH).

The invariance property (9) was seen to be verified in observational three dimensional galaxy catalogs (Alimi et al. 1990, Maurogordato et al. 1992). But the measurements were quite noisy, because of the smallness of these samples. In the much richer sets of points coming from $N$-body simulations, the relation (9) seems to be fulfilled with a great accuracy (BSD, BH) and the measured function $h$ has the expected asymptotic behaviors. However, testing the predictions of the scaling model on the CPDF is much more difficult than on the VPDF. Indeed, significant deviations from these predictions can be expected, even if the underlying distribution is perfectly scale invariant. Equation (9) is only valid in an asymptotic regime (never reached in practical cases) and for ideal sets of infinite volume. Because of the finiteness of the real samples size, the very large $N$ part of the count probability is unduly dominated by a few large clusters and therefore always presents a behavior incompatible with equation (9) (BSD, CBS). One consequence is that the direct measurement of the low-order moments is not realistic (Colombi & Bouchet 1991, CBS). Since one of the validity conditions of relation (9) (detailed in § 4) requires to be in the continuous limit, it can be expected that galaxy catalogs, dominated by discreteness, hardly reach this regime. In this case, one could thus wonder if the measurement of a function $h$ is really significant. This should be even worse for galaxy samples generated with a sparse sampling strategy (used to optimize the measurement of the two-body correlation function, Kaiser 1986). Also, the limit where a sample really disagrees or not with the scaling property has to be clearly defined, and dilution effects have to be carefully studied.

In this paper, we aim at first to list all the spurious effects that can bring systematic errors on the measurement of the VPDF and the CPDF. To be in the framework of the scaling model of BS, we analyze the functions $\hat{\sigma}(n,\ell)$ and $\hat{h}(N,n,\ell)$ rather than the VPDF and the CPDF. Our second objective is indeed to study the practicability of the very sophisticated formalism of BS. Hence, we will give the appropriate procedure to trustfully check the existence of a function $\sigma(y)$ and a function $h(x)$, and of course to determine these functions. Through the analysis of dilution effects, we will see that a sparse sampling strategy tends to reduce the scaling regime from which the more important informations on the VPDF and the on CPDF can be extracted.

Figure 1: A thin slice of width $L_{\rm box}/8$ extracted from our three samples of reference, i.e., the CDM sample $E$ (left panel), the spherical power-law cluster plunged into a poissonian noise (middle panel), and the fractal generated by a Rayleigh-Lévy random walk (right panel).

To do that, we shall use three samples as reference cases:

1. a CDM universe (left panel of Fig. 1) generated by a P$^3$M simulation involving 262 144 dark matter particles (Davis & Efstathiou 1988),

2. a cubical sample containing a power-law spherical cluster immersed in a poissonian noise (central panel of Fig. 1), that does not obey the scaling property (2),

3. a fractal generated by a Rayleigh-Lévy random walk (right panel of Fig. 1), which should, on the contrary, perfectly obey the scaling relation.

This paper is organized as follows: in § 2 we describe the above three samples. In § 3 we study the void probability and the function $\hat{\sigma}$. We try to evaluate the main misleading effects that can affect its measurement and to propose a procedure to determine fairly the function $\sigma$, if it exists. Section 4 is devoted to the function $\hat{h}$. We analyze in detail the predictions of BS for this function and we test the behavior of the system when it is diluted. We aim to see when a function $h$ might exist and what are the possible contamination effects that may hide it, such as finite volume effects. Conclusions are presented in § 5.

## 2 The samples

### 2.1 The CDM sample $E$

Our CDM sample (left panel of Fig. 1) was generated by Davis & Efstathiou (1988) with a P$^3$M code. It contains $N_{\rm par} = 262\,144$ matter particles, its physical size is $L_{\rm box} = 32 h^{-1}$ Mpc, and it was evolved until the variance computed in a sphere of radius 8 $h^{-1}$ Mpc reached $1/b$ with $b = 2.5$. The CPDF and the VPDF were computed by BSD for cubic cells of size $\ell$ in the range $-2.6 \leq \log_{10}(\ell/L_{\text{box}}) \leq -1.0$. These lower and upper limits were chosen in order to avoid respectively smoothing and "periodisation" effects. In the following we shall express all lengths in units of $L_{\text{box}}$.

BSD have measured the functions $\hat{\sigma}$ and $\hat{h}$ in the non-linear regime ($\overline{\xi}_2 > 1$). In particular, they found that $\hat{h}$ did indeed scale as a function $h$. They proposed a phenomenological fit for the function $h$, that we shall recall in § 4. They found that the transform (11) applied to this fit reproduced well the measured function $\hat{\sigma}$. However, all this work was done at constant number density $n = 262\ 144$. A full test of the scaling model needs $n$ to vary. Moreover, the current three-dimensional galaxy catalogs involve at most a few thousand objects. So to really test and understand the scaling model and its domains of use, it will be interesting to dilute our CDM sample that is known to exhibit the scaling properties predicted by BS.

## 2.2 The sample dominated by a cluster $E_{\text{c}}$

If the statistics is dominated by a single spherical and locally poissonian cluster with a power-law average profile, the low-order correlations do not obey the scaling relation (2). Indeed, following LSS, let us consider a cluster of radius $R$ for which the number density is

$$\begin{aligned} n(r) &= Ar^{-\alpha}, & r &\leq R, \\ n(r) &= 0, & r &> R, \quad 1.5 < \alpha < 3. \end{aligned} \tag{12}$$

According to LSS, when $r \ll R$, the correlation function $\xi$ verifies $\xi_2 \propto r^{3-2\alpha}$ and the low-order correlations scale as (Peebles & Groth 1975)

$$\frac{\xi_3}{\xi_2^2} \propto r^{\alpha-3}, \quad \frac{\xi_4}{\xi_2^3} \propto r^{2(\alpha-3)}. \tag{13}$$

So the scaling relation is not obeyed (because $\alpha \neq 3$). We have synthesized such a cluster, but for reasons of normalization and in order to get a more realistic $P_N(\ell)$, we have immersed it into a white noise: the density profile is then given by

$$\begin{aligned} n(r) &= Ar^{-\alpha}, & r &\leq R, \\ n(r) &= AR^{-\alpha}, & r &> R. \end{aligned} \tag{14}$$

The cluster is located at the center of the sample, which is cubical, of size $L_{\text{box}} \equiv 1$. The normalization $<n(r)> = n$ implies

$$AR^{-\alpha}\left(1 + \frac{4\pi}{3}\frac{\alpha}{3-\alpha}R^3\right) = n. \tag{15}$$

The calculation of the two-body correlation function gives, when $r \ll R$,

$$\xi_2(r) \sim 2\pi I A^2 n^{-2} r^{3-2\alpha}, \tag{16}$$

with

$$I = \int_0^\infty x^{2-\alpha} dx \int_{-1}^1 d\mu (x^2 + 1 + 2x\mu)^{-\alpha/2}. \tag{17}$$

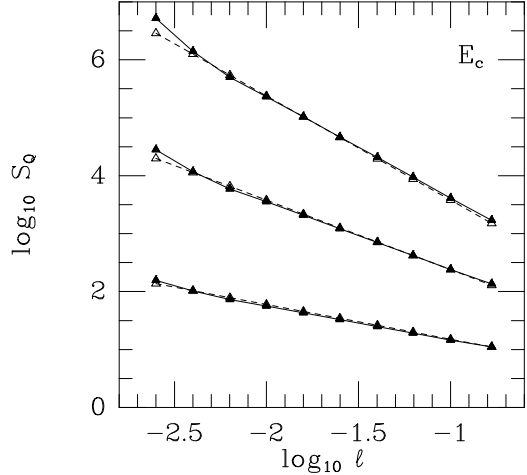

Figure 2: Quantities $S_Q \equiv \overline{\xi}_Q/\overline{\xi}_2^{Q-1}$ versus scale in logarithmic coordinates as measured in $E_c$ for $Q = 3, 4, 5$. $S_Q$ increases with $Q$. The dashed lines are power-laws of index $(Q-2)(\alpha-3)$ (eqs. [12], [13]).

Once $\alpha = (3+\gamma)/2$ is fixed, the choice of the number $nL_{\text{box}}^3$ of objects in the sample and of the correlation length $r_0$ determines $A$ and $R$. We have taken

$$\gamma = 1.8, \quad r_0 = 0.087, \quad n = 32768, \tag{18}$$

which leads to $\alpha = 2.4$, $A = 1.7 \; 10^{-2} n$ and $R = 0.19$.

The corresponding set $E_c$ is displayed in Fig. 1 (central panel). Since the white noise part of $E_c$ is not correlated, expressions (13) are still valid. This is illustrated by Fig. 2, which displays quantities $S_Q$, $Q = 3, 4, 5$ as functions of scale. The function $P_N(\ell)$, as measured on $E_c$ for cubic cells of size $\ell$ in the non-linear regime ($\overline{\xi}_2 \geq 1$) for $-2.6 \leq \log_{10}(\ell) \leq -0.8$, is given by Fig. 3. At fixed scale, it presents a bump at the vicinity of its maximum, corresponding to the poissonian noise that globally dominates the statistics (dashed curves), followed at larger $N$ by a power-law and a second bump above which $P_N(\ell)$ vanishes. The large $N$ part of the CPDF is dominated by the cluster statistics and can be easily evaluated (CBS). In particular, one can show that the power-law part of $P_N(\ell)$ verifies $P_N(\ell) \propto N^{-3/\alpha-1}$.

## 2.3 The Rayleigh-Levy fractal $F$

To have a trustable element of comparison, we have synthesized a sample $F$ which obeys the scaling property (2). It is a fractal involving 32 768 points generated by a Rayleigh-Lévy random-walk (see, e.g., Mandelbrot 1975, 1982, Peebles 1980, Colombi et al. 1992). Starting from a random point in the sample, the next point is chosen at random direction and at a distance $r$ with a probability

$$\begin{array}{ll} p(r > \ell) = (\ell_{\text{p}}/\ell)^\epsilon, & \ell \geq \ell_{\text{p}}, \\ p(r > \ell) = 1, & \ell < \ell_{\text{p}}. \end{array} \tag{19}$$

The process is repeated taking the new point as reference. To obtain the same two-body correlation function as in $E_c$, we chose $\epsilon = 3 - \gamma = 1.2$, $\ell_{\text{p}} = 1.24 \; 10^{-3}$. We refer to LSS (pages 245, 248) or

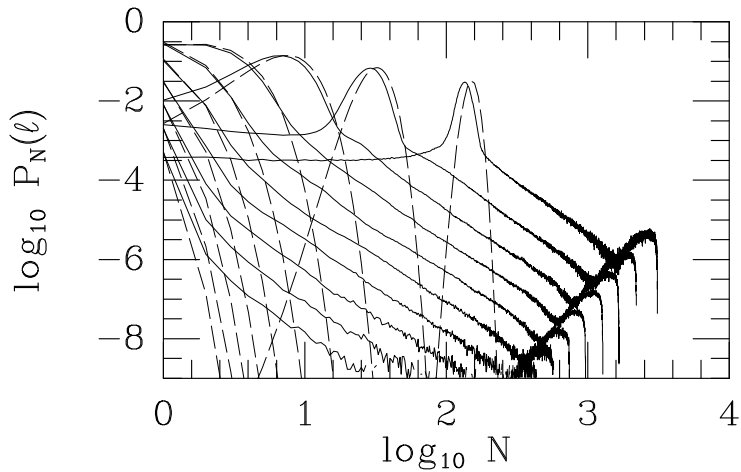

Figure 3: Logarithm of the count probability $P_N(\ell)$ measured on $E_c$ as a function of $\log_{10} N$, for various values of scale. The dashed lines correspond to the Poisson model.

Appendix A to have the expression of $\xi_2(r)$ in terms of $\epsilon$, $n$ and $\ell_p$. The finiteness of $\xi_2$ is insured by the fact that our sample is contained in a cube of size $L_{\text{box}}$ and involves a finite number of objects (so $F$ is not actually a real fractal). Figure 1 (right panel) shows a slice extracted from $F$ of width $L_{\text{box}}/8$. In Appendix A, we show, generalizing a calculation of Peebles for the low-order correlations (LSS, page 248), that this sample obeys the scaling relation (see also Hamilton & Gott 1988). We compute its function $\sigma$ and find the approximate expression

$$\sigma(y) \simeq (1 + y/2)^{-1}. \tag{20}$$

The function $h$ associated to equation (20) is

$$h(x) \simeq 4 \exp(-2x). \tag{21}$$

As for $E_c$, we have measured the count probability for cubic cells of size $\ell$ in the scale range $-2.6 \leq \log_{10}(\ell) \leq -0.8$. This sample is especially interesting, since we can exactly test on it the predictions of the scaling model. We are able to see if they are practically verified, despite the possible misleading effects we describe in the following. With this sample and $E_c$, we have two extreme reference cases to decide if a function $\sigma$ or a function $h$ exists or not.

## 3  The void probability (VPDF)

Strictly speaking, the notion of VPDF is meaningless for a continuous matter density field. Nevertheless, we can consider each matter particle of our CDM sample as a galaxy, as we shall do in the following. Then, one can measure the quantity $\hat{\sigma}(n, \ell)$ defined in introduction as a function of $N_c$ for various number densities $n$ to have an idea of the behavior of $P_0(n, \ell)$. For example, the first step is to see if the sample deviates from a pure poissonian distribution, for which $\hat{\sigma} \equiv 1$. Moreover, in this system of coordinates, if the scaling relation (2) applies, $\hat{\sigma}$ should scale as a single function $\sigma(N_c)$. One may however argue that a large class of samples may exhibit a function $\sigma(N_c)$, even if they do not obey the scaling property. The two samples $E_\mathrm{c}$ and $F$ we have generated are here to specifically address this problem. Some artificial deviations from the scaling relation may be induced by misleading effects, such as "grid effects" (taking place only in numerical simulations) and finite sample effects, that are studied in detail in Appendix B. We give the main results in § 3.4. The scaling relation may also work only in a finite range of scales, which does not exclude the existence of a function $\sigma(N_\mathrm{c})$ if only this regime is taken into account.

This section is thus organized as follows: in § 3.1, we recall some aspects of the formalism of BS. Section 3.2 studies and compares two ways of randomly diluting a sample, an analytical one and an experimental one. In § 3.3, we measure function $\hat{\sigma}$ in $E_\mathrm{c}$ and $F$. Section 3.4 deals with spurious effects. In § 3.5, we study the scaling behavior of the function $\hat{\sigma}$ measured in our CDM sample $E$.

## 3.1 Scaling model and VPDF

Here we assume that the scaling property (2) applies. It is argued in BS that the function $\sigma(N_\mathrm{c})$ should have a power-law behavior at large $N_\mathrm{c}$, i.e.,

$$\sigma(N_\mathrm{c}) \underset{N_\mathrm{c} \gg 1}{\propto} a N_\mathrm{c}^{-\omega} \tag{22}$$

with $0 \leq \omega \leq 1$ and $a > 0$. The measurement of $\sigma$ in the observed galaxy distribution is in agreement with equation (22). The CfA data provide $\omega \sim 0.5 \pm 0.1$ (Alimi et al. 1990) and the Southern Sky Redshift Survey (SSRS) data lead to $\omega \sim 0.7 \pm 0.1$ (Maurogordato et al. 1992). In universes of matter coming from numerical simulations, the value of $\omega$ seems to depend on initial conditions (see, e.g., BH). In our CDM sample, BSD have measured (for the distribution of matter particles, whereas the previous values concern the observed galaxy distribution, which is expected to be biased with respect to the matter distribution)

$$\omega \sim 0.4 \pm 0.05. \tag{23}$$

Let us define the number $N_\mathrm{v}$ by

$$P_0 \equiv \exp(-N_\mathrm{v}^{1-\omega}), \quad \omega < 1, \tag{24}$$

and $N_\mathrm{v} \equiv 0$ for $\omega = 1$. The number $N_\mathrm{v}$ can be considered as the typical number of objects in a cell located in an underdense region. The typical size $\ell_\mathrm{v}$ of a void naturally appears from the condition

$$N_\mathrm{v}(\ell_\mathrm{v}) = 1. \tag{25}$$

## 3.2 Dilution and VPDF

The function $\hat{\sigma} \equiv -\ln(P_0)/(n\ell^3)$ in principle depends on two variables, i.e., the average number density $n$ and the scale $\ell$. To check if the scaling relation is fulfilled, that is if $\hat{\sigma}(n, \ell) = \sigma(N_\mathrm{c})$, it is useful to display $\hat{\sigma}$ as a function of $N_\mathrm{c}$ for various number densities $n$, to cover all the dynamic range of possible values of $(n, \ell)$. Samples of various densities can be obtained by randomly extracting from the studied sample $E$ some subsamples $E_i^\mathrm{ran}$ ($i$ is the number of objects) with average number densities $n_i = n.i/N_\mathrm{par}$. These subsamples should have the same shape and the same volume than $E$, since each of them is formally a discrete realization of the same underlying density field than $E$, but with a number density smaller than in the set $E$. However, in observational data samples, this experimental dilution is often made by extracting several volume limited samples, so this condition is not verified, and some artifacts due to the variation of the subsamples physical size can contaminate the measurement.

On the other hand, if the function $P_N(n,\ell)$ of $E$ is known, this random dilution can also be done analytically (see Hamilton 1985, Hamilton et al. 1985). Indeed, one can easily show, using equation (4), that the CPDF $P_N(n,\ell)$, $N \geq 1$ can be obtained from the VPDF $P_0(n,\ell)$ through the following derivation (White 1979, BS)

$$P_N(n,\ell) = \frac{(-n)^N}{N!} \frac{\partial^N}{\partial n^N} P_0(n,\ell). \qquad (26)$$

This relation implies, by a simple Taylor expansion

$$P_0(n_i,\ell) = \sum_{N=0}^{\infty} \left(1 - \frac{n_i}{n}\right)^N P_N(n,\ell). \qquad (27)$$

One can moreover compute the count probability $P_N(n_i,\ell)$ of the subsample from $P_N(n,\ell)$ by applying equation (26) to equation (27):

$$P_N(n_i,\ell) = \sum_{K=N}^{\infty} C_N^K \left(\frac{n_i}{n}\right)^N \left(1 - \frac{n_i}{n}\right)^{K-N} P_K(n,\ell), \qquad (28)$$

with $C_N^K \equiv K!/[N!(K-N)!]$. The series expansion (27) converges at least for $n_i/n \leq 2$; the VPDF of a "subsample" denser than $E$ can therefore be predicted, with increasing error for increasing density. This error arises especially from the finiteness of the sample, which prevents function $P_N$ from being accurately calculated at large $N$.

Figure 4 shows $\hat{\sigma}$ as a function of $N_c$, as measured in our CDM simulation (triangles) and two randomly extracted subsamples $E_{32768}^{\rm ran}$ (squares) and $E_{1024}^{\rm ran}$ (pentagons). The dashed curves represent analytical dilutions on $E$ and on $E_{32768}^{\rm ran}$ with respective dilution factors $n/n_{32768} = 8$ and $n_{32768}/n_{1024} = 32$. They quite superpose to the squares and the pentagons, as expected. This measurement shows that, even if the number of points in the considered catalog is small, the direct determination of the VPDF in this set is accurate (in the available dynamic range). In other words, errors related to statistical poorness are small. Indeed, the analytical dilution (27) uses the statistics of the full set $E$ and should thus be more accurate than the experimental dilution.

One can similarly test equation (27) with $n_i/n = 2$. Figure 5 displays $\hat{\sigma}$ as a function of scale for $E_{1024}^{\rm ran}$ (pentagons) and $E_{512}^{\rm ran}$ (hexagons). (We use this system of coordinates to easily distinguish between the two subsamples). The dashed curve represents the function $\hat{\sigma}$ as it should be measured on $E_{1024}^{\rm ran}$, applying equation (27) to $E_{512}^{\rm ran}$ with $n_i/n = 2$. Again, the agreement between the measurement and the prediction is good, although the two subsamples are statistically poor.

Note that the available dynamic range where the function $\hat{\sigma}$ significantly differs from unity of course decreases with decreasing number density $n$. Now, the interesting properties of a scale invariant distribution are reached at large $N_c$ (eq. [22]) which of course argues against a sparse sampling strategy. We shall discuss again this problem in § 3.5.

## 3.3 How significant is a measurement of the function $\hat{\sigma}$?

Let us now see if at a first glance, the measurement of $\hat{\sigma}$ gives appropriate results for our reference sets $E_c$ and $F$. Figure 6 shows $\hat{\sigma}$ as a function of $N_c$, as measured on $E_c$ (left panel) and on $F$ (right panel), for several number densities $n_i = 2n, n, n/8, n/39.5, n/512, n/1024$ (that we also test on our CDM sample $E$), using the analytical procedure given by equation (27). The stars (on left panel) correspond to the cases $n_i < n$, the filled points to the direct measurement, and the open symbols correspond to the case $n_i = 2n$. As expected, the fractal exhibits an unique function $\sigma(N_c)$, that is quite identical to the theoretical expectation (20). The set $E_c$ does not scale as a function $\sigma(N_c)$ as well as $F$. Its function $\hat{\sigma}$ presents a plateau where $\hat{\sigma} \sim 1$ and a sudden cutoff at $N_c \gtrsim 10$. The plateau is not surprising, since $E_c$ is dominated by the poissonian distribution that surrounds the central cluster. The main difference with a pure poissonian sample is the existence of a large range of values of $N_c$, which is provided by the presence of the cluster. A poissonian sample, that is not correlated, has $N_c = 0$. The points that verify $P_0(\ell) \leq \ell^3$ are circled in left panel of Fig. 6. They precisely correspond to the sharp cutoff at large $N_c$. As we shall see in next section, this cutoff does not reflect the intrinsic properties of the underlying distribution, but some special properties of the largest void contained in this particular realization. The set $F$ has much larger empty regions, so none of the studied scales are susceptible to exhibit such an abnormal behavior. Therefore, with regard to $E_c$ and $F$, a careful measurement of function $\hat{\sigma}$ seems to really reflect the underlying behavior of the studied distribution.

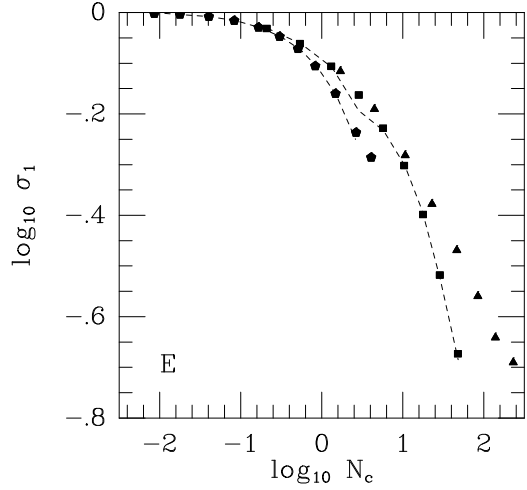

Figure 4: $\hat{\sigma} \equiv -\ln P_0/(n\ell^3)$ as a function of $N_c \equiv n\ell^3 \overline{\xi}_2$ in logarithmic coordinates. The triangles refer to our CDM sample $E$, the squares and the pentagons to the randomly extracted subsamples $E^{\rm ran}_{32768}$ and $E^{\rm ran}_{1024}$ respectively. The dashed curves correspond to an analytical dilution on $E$ and $E^{\rm ran}_{32768}$ by a factor 8 and a factor 32 respectively; they should superpose to the squares and the pentagons.

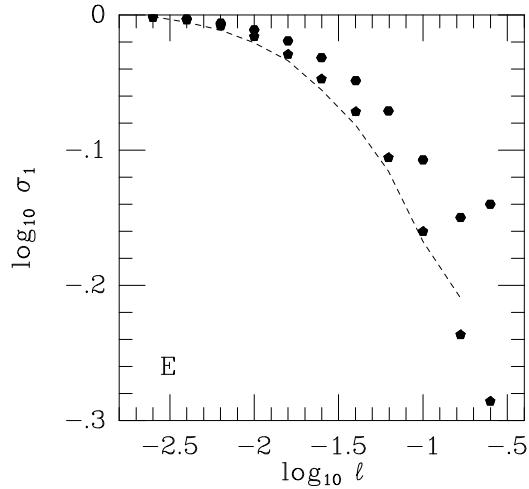

Figure 5: $\log_{10} \hat{\sigma}$ as a function of logarithm of scale in the very dilute regime. Pentagons correspond to $E^{\rm ran}_{1024}$ and hexagons to $E^{\rm ran}_{512}$. The dashed curve represents a virtual increase in number density by a factor 2 applied to $E^{\rm ran}_{512}$. It should superpose to the pentagons.

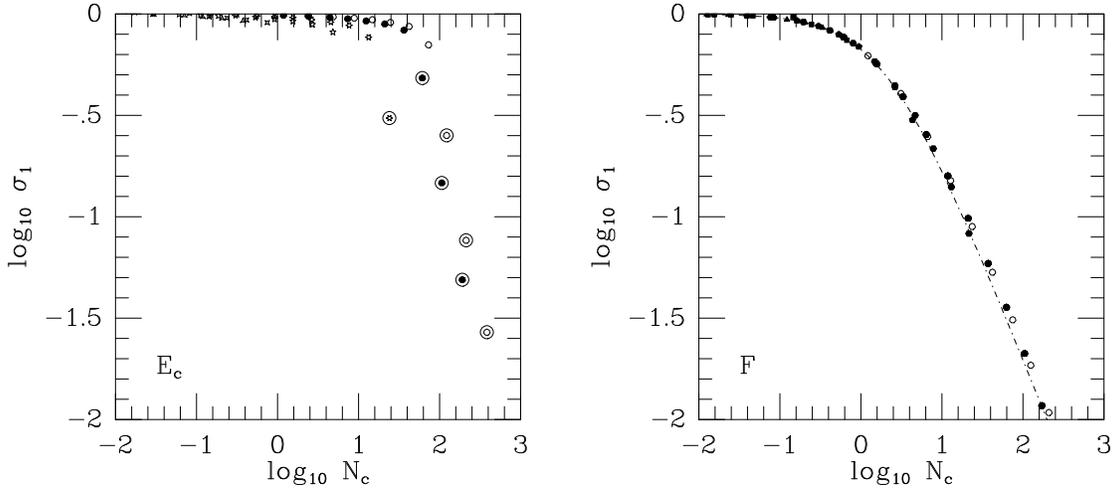

Figure 6: Logarithm of $\hat{\sigma}$ as a function of logarithm of $N_c$ for $E_c$ (left panel) and $F$ (right panel). The filled points correspond to the direct measurement, the stars on left panel to various analytical dilutions varying from a factor 8 to 1024, and the open symbols to a virtual multiplication of average number density by a factor 2. The circled points on left panel correspond to scales where $P_0(\ell) \leq \ell^3$. The dotted-dashed curve on right panel is the theoretical expectation.

### 3.4 Misleading effects

We now consider possible misleading effects on the VPDF. At first we discuss "grid effects" that may take place in numerical simulations. Then, we look at finite sample effects on the VPDF and on function $\hat{\sigma}$. This section ends with a discussion on a recent measurement of function $\hat{\sigma}$ by Vogeley et al. (1991) on the CfA catalog. We also see what happens to our CDM sample $E$.

### 3.4.1 Grid effects

Grid effects only exist in numerical simulations. They are linked to the fact that the information associated to the grid of initial particle positions in the simulation has not been completely destroyed in large underdense regions, where there is no shell-crossing. Their consequence, as shown in Appendix B.1, is that the VPDF is artificially underestimated at large scales. They are negligible when

$$P_0 \gtrsim 1/e, \tag{29}$$

or equivalently when $N_v \lesssim 1$ in the formalism of BS. The scales for which $P_0 < 1/e$ cannot be trustfully tested and have to be removed. This decreases the available dynamic range in which the scaling of the function $\hat{\sigma}$ scales as a function $\sigma$ can be tested. These effects do not exist in the galaxy catalogs.

### 3.4.2 Finite sample effects and the VPDF

Practically, the measurement of VPDF is done by randomly throwing a certain number $C_{\text{tot}}$ of cells of volume $\ell^3$ in the sample. It is then easy to show that the standard deviation on $P_0$ is related to $C_{\text{tot}}$ (see for instance Hamilton 1985) through the following expression

$$\left(\frac{\Delta P_0}{P_0}\right)^2 = \frac{1}{C_{\text{tot}}} \frac{1 - P_0}{P_0}. \tag{30}$$

But, as already discussed for example by Maurogordato & Lachièze-Rey (1986), since the sample is of finite volume, the number of statistically independent cells $C_{\text{tot}}$ is not arbitrarily large and may depend on average number density $n$ and on scale $\ell$.

For a pure poissonian distribution (not correlated), a natural guess is simply

$$C_{\text{tot}} = V_{\text{sample}}/\ell^3, \tag{31}$$

which roughly gives the number of "statistically independent" cells of volume $v = \ell^3$ contained in the sample. This estimation is valid for a moderately small VPDF (see Appendix B.2.1). In the case the VPDF is very small, a correction to equation (31) is needed and $C_{\text{tot}}$ is rather of the order of $0.07 V_{\text{sample}} \ell^{-3} (n\ell^3)^3$ for spherical cells (see Appendix B.2.1, B.2.2).

The case of a correlated set is more complicated. Indeed, some correlations at scales larger than the sample size are then likely to affect the measurement of the VPDF. It is not easy to see where they intervene in equation (30). Our aim here is to try to clarify this situation. However, we take a slightly different approach from the one used by previous authors, which was generally based on equation (30) (with a somewhat uncertain $C_{\text{tot}}$).

The technical issues are detailed in Appendix B.2, where we evaluate the error associated to finite volume effects on the VPDF, assuming that the hierarchical model applies, i.e., that the $Q$-body correlation function can be written as sum of products of $Q - 1$ terms in $\xi_2$. We think that the result can be reasonably generalized for any sample which does not disagree "too much" with the hierarchical model (an independent estimation using the top hat model leads to similar results, see Appendix B.2). The error on the VPDF is roughly approximated by

$$\left(\frac{\Delta P_0}{P_0}\right)^2 \simeq \Sigma_1 + \Sigma_2, \quad \Sigma_1 = \frac{\ell^3}{V_{\text{sample}}} \frac{1 - P_0}{P_0}, \quad \Sigma_2 = -2(n\ell^3)^2 \hat{\sigma}' \overline{\xi}_2(L_{\text{sample}}), \tag{32}$$

where $\hat{\sigma}' \leq 0$ stands for the partial derivative of function $\hat{\sigma}(N_c, n)$ with respect to $N_c$ and $\overline{\xi}_2(L_{\text{sample}})$ symbolizes the double integral of $\xi_2(\mathbf{r}_1, \mathbf{r}_2)/V_{\text{sample}}^2$ over the sample volume $V_{\text{sample}} \sim L_{\text{sample}}^3$.

The first right hand side term $\Sigma_1$ of equation (32) can easily be understood. It is equal to the right hand side of equation (30), with $C_{\text{tot}}$ given by equation (31). It therefore simply corresponds to the expected "poissonian" error due to the fact that the number of statistically independent cells of volume $v = \ell^3$ contained in the sample is finite. As it is the case for a pure poissonian sample, a correction to $\Sigma_1$ is needed when the VPDF is very small (see appendix B.2.1, B.2.2). For practical purposes, we neglect this correction. Indeed, when it has to be taken into account, the relative error $|\Delta P_0/P_0|$ is likely to be close to unity or larger. An other factor $\alpha$ depending on the way the system is clustered [i.e., on $\gamma$ and on the shape of function $\sigma(y)$] has also been neglected in the writing of $\Sigma_1$, since it is numerically seen to be of order unity within less than a magnitude (and it is of course exactly equal to unity in the poissonian case). The second right hand side term $\Sigma_2$ of equation (32) is brought by fluctuations of the underlying density field at wavelengths larger than the sample size, which lead supplementary correlations.

We see that, whatever the value of the second right hand side term, there is a scale $\ell_{\text{cut}}(n)$ above which the measurement of the VPDF is not statistically significant. At this scale, which is defined by

$$P_0(n, \ell_{\text{cut}}) \frac{V_{\text{sample}}}{\ell_{\text{cut}}^3} \equiv 1, \tag{33}$$

there is typically only one independent empty cell. If $\ell \gtrsim \ell_{\text{cut}}$, the VPDF is dominated by the largest void of the sample. In a set of infinite volume and with an infinite number of objects (but a finite number density $n$), we should have an infinite distribution of voids of arbitrary size. This is not the case for a realization of finite volume, in which the size of the largest void is necessarily smaller than the size of the sample.

If the formalism of BS applies, $\Sigma_2$ can be rewritten, for $N_c \gg 1$,

$$\Sigma_2 = 2\omega n \ell^3 \sigma(N_c) \frac{\overline{\xi}_2(L_{\text{sample}})}{\overline{\xi}_2} = -2\omega \ln(P_0) \frac{\overline{\xi}_2(L_{\text{sample}})}{\overline{\xi}_2}. \tag{34}$$

One can thus expect this term to be in general small for moderately small VPDF. The poissonian error $\Sigma_1$ also becomes rapidly very small if $\ell$ gets small as compared with $\ell_{\text{cut}}$.

### 3.4.3 Finite sample effects and function $\hat{\sigma}$

We now wish to estimate the finite sample error on the quantity $\hat{\sigma} \equiv -\ln(P_0)/(n\ell^3)$. Assuming that the uncertainties $\Delta P_0/P_0$ and $\Delta n/n$ are small and that the indicators $\widetilde{P}_0$ and $\widetilde{n}$ are statistically independent, we might use the standard errors propagation formula and write

$$\left(\frac{\Delta \hat{\sigma}}{\hat{\sigma}}\right)^2 \simeq \frac{1}{(n\ell^3 \hat{\sigma})^2} \left(\frac{\Delta P_0}{P_0}\right)^2 + \left(\frac{\Delta n}{n}\right)^2, \tag{35}$$

where

$$\left(\frac{\Delta n}{n}\right)^2 = \frac{1}{n V_{\text{sample}}} + \overline{\xi}_2(L_{\text{sample}}). \tag{36}$$

However, the VPDF and the number density $n$ are strongly correlated. If $n$ increases, then $P_0$ decreases. In other words, equation (35) is likely to overestimate the real uncertainty on $\hat{\sigma}$. It is therefore certainly more realistic to use the following estimate

$$\frac{\Delta \hat{\sigma}}{\hat{\sigma}} \simeq \left| \frac{1}{n\ell^3 \hat{\sigma}} \frac{\Delta P_0}{P_0} - \frac{\Delta n}{n} \right|. \tag{37}$$

This writing implies that, as expected, $\Delta \hat{\sigma}/\hat{\sigma} \simeq 0$ when $P_0$ is close to unity (or, equivalently, when $n\ell^3 \hat{\sigma} \sim 0$), which was not the case with the more approximate evaluation (35).

When $\ell < \ell_{\text{cut}}$, it is easy to see that

$$\frac{\Delta \hat{\sigma}}{\hat{\sigma}} \lesssim \frac{\Delta n}{n}. \tag{38}$$

This error should therefore be small if the size of the studied sample is large in comparison with its correlation length and if $nV_{\text{sample}} \gg 1$, which is the case in current observational data volume limited subsamples, that involve at least a few hundred objects.

In fact, when one displays $\hat{\sigma}$ as a function of $N_c$, a supplementary (and quite larger) error is brought by the uncertainty on the determination of $N_c = n\ell^3 \overline{\xi}_2$ (see Hamilton 1993 for a detailed study of the errors on the two-body correlation function). Indeed, the function $\overline{\xi}_2$ is very sensitive to finite volume effects and is likely to be quite underestimated at large scales (CBS). The consequence is that the measured $N_c$ reaches its maximum much sooner than expected. In Appendix B.3, we estimate the error linked to the finiteness of the sampled volume on function $\xi_2$. For a set obeying to the scaling property, one obtains

$$\left( \frac{\Delta \overline{\xi}_2}{\overline{\xi}_2} \right)^2 \simeq \frac{S_4}{4} \overline{\xi}_2(L_{\text{sample}}). \tag{39}$$

This expression is valid when the considered scale $\ell$ is small enough compared to the sample size $L_{\text{sample}}$ (see Appendix B.3). (Note that the relative error on $N_c$ is slightly larger, since one has to take also into account the error on the average number density $n$, which is small in comparison with the uncertainty associated to the two-body correlation). In addition, we must not forget the error on $\overline{\xi}_2$ associated to the sample discreteness (i.e., that the number of sampled pairs is finite, see, e.g., Peebles 1973, LSS, p. 189 and Hamilton 1993), that we do not estimate here.

### 3.4.4 Comments

When Vogeley et al. (1991) measure the function $\hat{\sigma}$ in the CfA catalog, they conclude that there is a deviation from the scaling relation at $\ell \geq 10h^{-1}$ Mpc. But at such a large scale, the measured number $N_c$ is, as discussed above, likely to be underestimated. Moreover, the measured function $\hat{\sigma}$ becomes an increasing function of $\ell$ because the sample is dominated by its largest void (Appendix B.2.2). In this case, the measurement of $P_0$ is not statistically significant, so no conclusion can be reached about the scaling behavior of the underlying galaxy distribution at such scales, not because the scaling does not exist, but because the sample is too small. This point has also been discussed by Maurogordato et al. (1992).

With regard to our CDM sample $E$, we know from CBS that finite volume effects are negligible on $N_c$ in the scaling range measured. Such effects should be even less significant on function $\hat{\sigma}$. Actually, grid effects are here more important than finite volume effects, and the condition $P_0 \geq 1/e$ is stronger than the condition $\ell < \ell_{\text{cut}}$.

## 3.5 Extracting the possibly scaling part of the VPDF

In Fig. 7, we give $\hat{\sigma}$ as a function of $N_c$ for $E$, and several subsamples extracted at random (with 32768, 6634, 1024 and 512 objects). Open triangles represent a virtual increase of the average number density of $E$ by a factor 2. The points for which $P_0 < 1/e$ have been removed, although grid effects should be less significant for randomly diluted subsamples, particularly if the dilution factor is large. Indeed, the information linked to a grid is partially destroyed by random dilution. The deviation from the scaling, even if it still exists, is much less apparent than in Fig. 4. But as noticed above, removing the scales for which $P_0 \leq 1/e$ decreases the available dynamic range and improves the apparent existence of a function $\sigma(N_c)$.

We also know from the measurement of the low-order averaged matter correlations (CBS) in this sample that there should be a deviation from the scaling relation when $\log_{10}\ell > -1.6$, because of the transition around $\ell_0$ between the highly non-linear regime and the weakly non-linear regime, that exhibit here different values of $S_Q$ for $Q \leq 5$ (see also § 4.2.1 hereafter). To measure the function $\sigma$, if it exists, one must at first select the scale range $L_{\text{scaling}}$ where the scaling relation is expected to apply. The measurement of the low-order correlations provides at least the upper bound of $L_{\text{scaling}}$ (the lower bound of $L_{\text{scaling}}$ is difficult to evaluate in numerical samples, because at very small scales, the low-order correlations are affected by discreteness and also by numerical errors linked to the smoothing of the forces in the program solving the equations of motion). This measurement must be done with care, since even the low-order correlations are quite sensitive to finite sample effects (CBS). In Fig. 7, the points that verify $\log_{10}\ell \leq -1.6$ are circled. They unambiguously define a single function $\sigma(N_c)$. This proves that the measured low-order statistics is here in agreement with the measured high-order statistics (function $\hat{\sigma}$ is related to the behavior of the averaged correlations at any order), as far as the scaling relation is concerned.

The dotted-dashed curve in Fig. 7 is the function $\sigma(N_c)$ predicted by the transformation (11) applied to the function $h(x)$ measured on $E$ by BSD (see § 4.1). It is in very good agreement with the measurement (circled points). This proves that with careful measurements, the predictions of BS are obeyed in a great detail, even on relations between indicators that describe very different regions of the studied sample: the function $\sigma(N_c)$ tests underdense regions, whereas the function $h(x)$ tests overdense places.

Note however that the very dilute subsamples ($n \leq 6634$) provide at most $N_c \sim 2$ when $\log_{10}(\ell) \leq -1.6$: at such a $N_c$, the asymptotic power-law behavior (22) of $\sigma$ is not reached, although $\sigma$ already significantly deviates from the Poisson expectation ($\sigma = 1$). With regards to the perfectly scale invariant fractal $F$, for $n = 6634$, we have $N_c(\ell_0) \sim 40$. With smaller values of $N_c$ (especially smaller than 10), we would still miss the large-$N_c$ power-law behavior of $\sigma$. This suggests a lower bound on $N_c$ to determine the power-law behavior of $\sigma$: it has at least to be larger than a few tens at the correlation length, that is $n\ell_0^3 \gtrsim 30-40$, or, in terms of galaxy number density ($\ell_0 \simeq 2.4 r_0 \simeq 12 h^{-1}$ Mpc)

$$n \gtrsim 0.02 - 0.03 h^3 \text{ Mpc}^{-3}. \tag{40}$$

This average number density is hardly reached by current observational volume limited subsamples extracted from data samples, in which $n$ is at most of order $0.01 h^3$ Mpc$^{-3}$, or equivalently $N_c(\ell_0) \sim 15$. This, of course, argues against any sparse sampling strategy. Indeed, since the VPDF is not very sensitive to finite volume effects, it is more important to have a small but as complete as possible catalog than a large dilute sample to measure the function $\hat{\sigma}$.

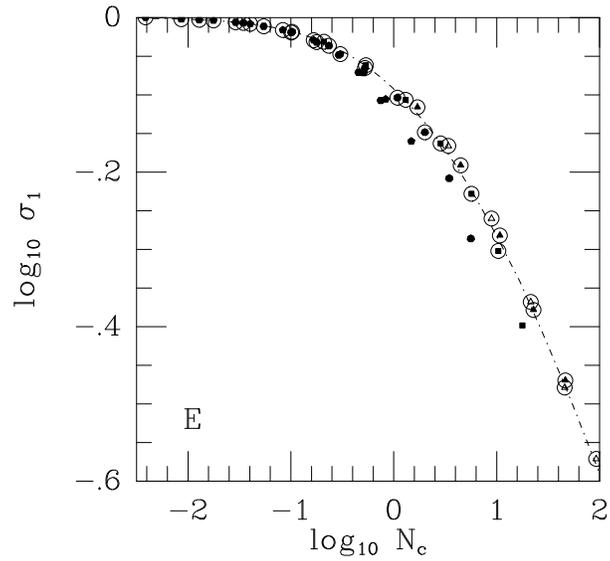

Figure 7: $\hat{\sigma}$ as a function of $N_c$ in logarithmic coordinates, as measured in our CDM sample $E$, and several subsamples extracted at random. The open triangles represent the measurement on a virtual sample twice denser than $E$, using the analytical prescription (26). The points for which $P_0 < 1/e$ have been removed. The circled points verify $\log_{10} \ell \leq -1.6$. In this scale range, the low-order correlations obey the scaling relation. The dotted-dashed curve is the analytical transformation (11) applied to the function $h$ BSD have measured on $E$.

# 4 The count probability distribution function (CPDF)

In the scaling model framework, it is useful to measure the quantity $\hat{h}(N, n, \ell) \equiv N_c \overline{\xi}_2 P_N(n, \ell)$ as a function of $x = N/N_c$, for various values of $(n, \ell)$. Indeed, when the scaling relation (2) applies, $\hat{h}$ behaves like an universal function function $h(x)$ in an asymptotic domain $D_h$. Practically, however, this asymptotic regime is never completely reached and one can expect some contamination effects to modify the behavior of function $\hat{h}(N, n, \ell)$ so that it does not scale exactly as a function $h(x)$. For example, the high $N$ part of the function $P_N(n, \ell)$ presents increasingly large irregularities as $N$ grows and a brutal cutoff at finite $N$, because of the sample volume finiteness, as shown by CBS.

Now, the CPDF is usually studied through its reduced moments, e.g., the averaged correlation functions $\overline{\xi}_Q$. Unfortunately, the measurement of such functions is very sensitive to the finiteness of the sampled volume. CBS have studied in detail such finite volume effects on the CPDF, and concluded they could lead to a systematic, strong, underestimation of the real low-order moments of the CPDF with direct measurements. They proposed a method to correct for such defects, or at least to estimate fair errorbars. We shall recall below their main results.

Once all spurious effects are known, one can wonder if the apparent existence of a function $h$ is not more or less systematic whatever the studied sample. In such a case, the formalism of BS would be useless. Once we will be convinced that this is not the case, we shall be able to see to what extent it is possible to detect a function $h(x)$ in the observed galaxy distribution.

This section is thus organized as follows: in § 4.1, we give $D_h$ and the main properties of function $h$, that have been computed by BS. We study contamination effects brought by discreteness and finite sample effects in § 4.2 (where we recall the main results of CBS). In § 4.3, we measure function $\hat{h}$ on the three reference samples $E$, $E_c$ and $F$. By randomly diluting our CDM sample $E$, we also see what occurs when the number density of the sample becomes comparable to which is reached in current three-dimensional observational data samples. We then try to determine a criterion on the galaxy number density to measure the interesting scaling behavior of function $\hat{h}$, as we did for the VPDF.

## 4.1 Some aspects of the scaling model linked to overdense regions

Here we recall some predictions of BS based on the scaling relation (2).

The invariance property (9) is expected to be reached in the scale and number domain $D_h$ given by

$$D_h \equiv \{\ell_c \ll \ell \ll \ell_0, \quad N \gg 1, \quad N \gg N_v\}. \tag{41}$$

$\ell_0 \sim 2.4 r_0$ is the (pseudo-)correlation length for which the averaged two-body correlation is unity:

$$\overline{\xi}_2(\ell_0) \equiv 1. \tag{42}$$

The scale $\ell_c$ is defined by

$$N_c(\ell_c) \equiv 1. \tag{43}$$

It is the typical distance between two objects in a cluster. When $\ell \ll \ell_c$, the system becomes quasi-poissonian, i.e., $\sigma(N_c) \sim 1$. The condition $N \gg 1$ is linked to discreteness effects. If $N_v \gg 1$, the CPDF presents at low $N$ ($N \ll N_v$) an exponential cutoff which scales as a function $g$. The latter is completely determined once $\omega$ is known (eq. [22]), so its main interest is to help for the measurement of $\omega$, in the case the scaling relation applies.

The function $h$ should behave at small $x$ as a power-law:

$$h(x) \sim x^{\omega-2}, \qquad x \ll 1, \tag{44}$$

and it presents at large $x$ an exponential cutoff

$$h(x) \sim \exp(-|y_s|x), \qquad x \gg 1. \tag{45}$$

It must also obey the normalization conditions (see, e.g., BS)

$$\int_0^\infty x h(x) dx \equiv 1, \qquad \int_0^\infty x^2 h(x) dx \equiv 1. \tag{46}$$

The CPDF measured in our CDM sample was found by BSD to scale properly and determine a unique function $h$ when $(N, \ell) \, \epsilon \, D_h$ (see also § 4.3 hereafter). BSD have fitted $h$ with the following phenomenological form

$$h_{\text{fit}}(x) = a \frac{(1-\omega)}{\Gamma(\omega)} \frac{x^{\omega-2} e^{-|y_s|x}}{(1+bx)^c}, \tag{47}$$

which, of course, obeys the normalizations (46), follows the small-$x$ power-law behavior (44) and has an exponential cutoff (45) at large $x$ with $|y_s| \sim 0.125$. The parameters $a \sim 1.8$ and $\omega \sim 0.4$ are determined by the measurement of $\sigma(N_c)$ at large $N_c$ (see eq. [22], and also by the measurement of function $g$, see, e.g., BSD). The parameter $|y_s|$ is determined by the measurement of function $h$ at large $x$. The remaining parameters $b \sim 3.6$ and $c \sim 0.8$ are determined by the two constraints (46).

## 4.2 Misleading effects on the CPDF and on function $\hat{h}$

### 4.2.1 CPDF and finite volume effects

The CPDF, as it is measured in universes of matter coming from $N$-body simulations (see, e.g., BSD, BH, CBS) or in the galaxy distribution (see, e.g., Alimi et al. 1990), is seen to present, in the non linear regime, an exponential tail at large $N$, with irregularities more and more pronounced as $N$ increases and a brutal cutoff at a finite $N = N_{\text{max}}$, above which $P_N = 0$. This is illustrated by figure 8, which displays the quantity $\log_{10} N^2 P_N$ measured in our CDM sample $E$ as a function of $N$. In current observed three-dimensional galaxy catalogs, for example the SSRS catalog analysed by Maurogordato et al. (1992), these irregularities are so pronounced that the exponential tail of the CPDF is difficult to detect. CBS have shown that such irregularities are due to the fact that, at large $N$, the CPDF is dominated by a few large clusters in the sample, each bump corresponding to a cluster. The last bump presented by each curve (for $-2.0 \lesssim \log_{10} \ell \lesssim -1.6$) on Fig. 8 corresponds to the largest cluster in our CDM sample (see CBS for a detailed modelization and discussion). Such irregularities and the brutal cutoff at $N_{\text{max}}$ are of course spurious. In a larger sample, the cutoff and the bumps would appear further down the large-$N$ tail of the CPDF: the bumps in the smaller sample would now be smoothed by statistical averaging. Indeed, in an infinite volume, one can find an infinite number of clusters of any sizes. The exponential tail shown up by the CPDF at large $N$ should thus be extended to infinity. Of course, the range of the exponential tail that brings a substantial contribution to any statistics depends on that statistics, and it does not need to be known with precision at arbitrarily large $N$. It can be modeled in the following way

$$P_N(\ell) \sim A(\ell) N^{\eta(\ell)} e^{-\beta(\ell)N}, \tag{48}$$

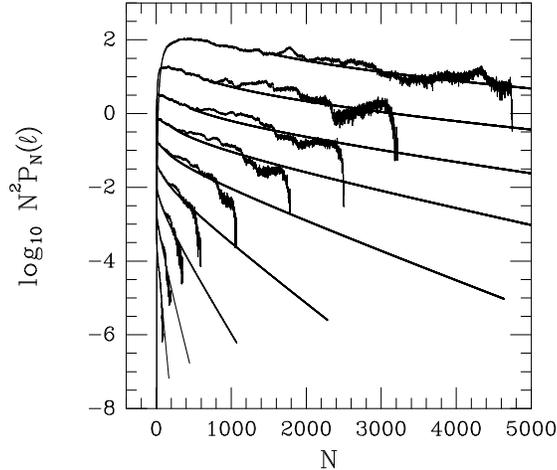

Figure 8: Logarithm of the quantity $N^2 P_N(\ell)$, measured in our CDM sample, as a function of $N$. Each curve corresponds to a given choice of scale. The scale decreases from the top curve (for which we have $\log_{10} \ell = -1.0$) to the bottom curve (for which we have $\log_{10} \ell = -2.6$) with a logarithmic step $\Delta \log_{10} \ell = 0.2$. The smooth lines are the analytical fit given by eq. (48). This figure is extracted from CBS.

with
$$\beta(\ell) = |y_{\rm s}(\ell)|/N_{\rm c}(\ell). \tag{49}$$

The functions $\eta(\ell)$ and $|y_{\rm s}(\ell)|$ are expected to slowly vary with scale (and to be constant if the scaling relation applies, see § 4.1), and the quantity $A(\ell)$ is a normalization factor (the best fit values of $\eta(\ell)$ and $|y_{\rm s}(\ell)|$ can be found in CBS).

Let us now turn to the $Q$-body averaged correlation functions $\overline{\xi}_Q$. When the order $Q$ increases, the weight given to high density regions is larger and larger and the defects described above, particularly the cutoff of the CPDF at $N_{\rm max}$, have stronger and stronger influence on the measurement of $\overline{\xi}_Q$. Such measurement would be more realistic if one extended to infinity the exponential tail (48). To illustrate this point, figure 9 displays in logarithmic coordinates the quantities $S_Q \equiv \overline{\xi}_Q/\overline{\xi}_2^{Q-1}$, $3 \leq Q \leq 5$, as functions of $\overline{\xi}_2$, measured in $E$ and in a larger CDM realization $E_{\rm L}$ (with a box size $L_{\rm box} = 90 h^{-1}$ Mpc). The underlying statistics in $E$ and $E_{\rm L}$ is theoretically exactly the same, but the set $E_{\rm L}$ is of much larger size than $E$ compared to the correlation length and is thus expected to be much less contaminated by finite volume effects. The solid curves with triangles and squares correspond respectively to direct measurements of functions $S_Q$ in the sets $E$ and $E_{\rm L}$. They undoubtly disagree between each other. But once finite volume effect in $E$ have been corrected by extending the exponential tail at large $N$ of the CPDF to infinity, one obtains a much better agreement (dashed curves). The same procedure applied to $E_{\rm L}$ does not significantly change the results. The above discussion shows that the method proposed by CBS to correct for finite volume effects is efficient (more tests are made in CBS to test the viability and the uncertainties of the method). Note that, in the sample $E$, the plateau exhibited by $S_Q$ in the highly non linear regime ($\overline{\xi}_2 \gg 1$) is hidden by finite volume effects, before correction.

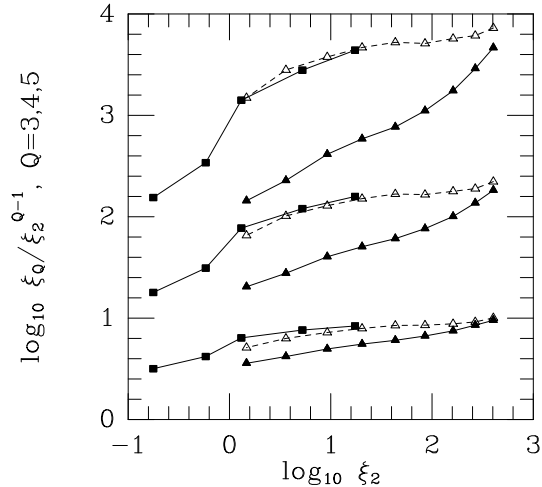

Figure 9: Logarithm of the quantity $S_Q \equiv \overline{\xi}_Q/\overline{\xi}_2^{Q-1}$, $3 \leq Q \leq 5$, measured in our CDM sample $E$ as a function of $\overline{\xi}_2$. The order $Q$ increases with $S_Q$. The filled triangles refer to $E$, the open triangles correspond to the same quantity, but corrected for finite volume effects and the squares refer to the larger CDM sample $E_{\rm L}$. Consistency between $E$ and $E_{\rm L}$ is only insured after correction. This figure is extracted from CBS.

### 4.2.2 Misleading effects on function $\hat{h}$

Practically, the regime defined by equation (41) is never reached, even in very large samples such as our CDM sample, that involve more than $2.10^5$ points. In three dimensional galaxy catalogues, that contain at most a few thousand objects, $\ell_{\rm c}$ is close to $\ell_0$ (at best of order $\ell_0/10$) and the constraint $\ell_{\rm c} \ll \ell \ll \ell_0$ is not verified. Then, the asymptotic regime (9) is hardly reached and the formalism of BS may be useless in this case.

Hence, let us assume that we have a sample of finite volume at our disposal, involving a finite number of objects and being a realization of a scale invariant underlying distribution. We evaluate here deviations from the invariance property (9) brought by the finiteness of the sampled volume and related to the fact that instead of equation (41), one practically has

$$\ell_{\rm c} \lesssim \ell \lesssim \ell_0, \quad N \gtrsim 1, \quad N \gtrsim N_{\rm v}. \tag{50}$$

To study separately the various misleading effects that can hide function $h$, we formally write (although the two terms $A_{\rm dis}$ and $A_{\rm fin}$ explained hereafter can be correlated each other)

$$\hat{h}(N, n, \ell) = U(N, \ell, n).h(x), \tag{51}$$

with

$$U(N, \ell, n) = [1 + A_{\rm dis}(N, n, \ell)].[1 + A_{\rm fin}(N, \ell)]. \tag{52}$$

The first factor of equation (52), $1 + A_{\rm dis}$, accounts for discreteness effects and for the deviation from the scaling behavior of the system in underdense regions ($N \sim N_{\rm v}$). It has been evaluated by BSD, assuming that function $\sigma(N_{\rm c})$ has reached its expected power-law behavior (22) at large $N_{\rm c}$ (with the notations of BSD, we have $1 + A_{\rm dis} = 1/a_N$). Note that the quantity $A_{\rm dis}$ computed by BSD depends on the parameter $\omega$ defined in equation (22). In appendix C.1, we study $A_{\rm dis}$ in terms of moments of the CPDF (without making any assumption on function $\sigma$). For example, in the vicinity of $N \sim N_{\rm c}$, the quantity $1 + A_{\rm dis}$ should be roughly of the same order as

$$1 + A_{\rm dis}(N, n, \ell) \sim 1 + 1/N_{\rm c} + 1/\overline{\xi}_2, \quad N \sim N_{\rm c}. \tag{53}$$

This estimate is only very approximate, because it does not take into account the complex behavior of the number $A_{\rm dis}$ as a function of $N$. Of course, in the limit $N \gg 1$, $N \gg N_{\rm v}$ and $\ell_{\rm c} \ll \ell \ll \ell_0$, $A_{\rm dis}$ vanishes.

The second factor, $1 + A_{\rm fin}$, is related to finite volume effects, that are widely discussed in previous section. Because of the finiteness of the sampled volume, the CPDF presents irregularities at large $N$ followed by a sharp cutoff at $N = N_{\rm max}(n, \ell)$, instead of the smooth exponential tail predicted at large $x$ for function $h$ (eq. [45]). These effects have been carefully studied by CBS. A cutoff at large $N$ on the CPDF changes the overall normalizations. Indeed, $1 + A_{\rm fin}$ is given, in the vicinity of $N_{\rm c}$, by

$$1 + A_{\rm fin}(N, \ell) \sim 1 \bigg/ \int_0^{N_{\rm max}/N_{\rm c}} x^2 h(x) dx \quad > 1. \tag{54}$$

This result is valid only if the two-body correlation is not affected by finite volume effects when $\ell \lesssim \ell_0$, which should be the case of any reasonable sample. When the sample size gets larger, the ratio $N_{\rm max}/N_{\rm c}$ increases. In the limit of an infinitely large sample, $1 + A_{\rm fin}$ goes of course to unity.

## 4.3  Is there a function $h$ or not?

We can see, from results of previous section, that the measurement of function $\hat{h}$, in the framework of the scaling model, is quite delicate. When the above defects are important, it may be very difficult to distinguish between a sample which obeys the scaling property and a sample which does not. Here, we measure function $\hat{h}$ on the two reference samples $E_{\rm c}$, $F$, and on the more realistic set $E$ and its various subsamples obtained by random dilution. Our aim is to prove that a careful measurement of function $\hat{h}$ allows one to successfully test the predictions of BS, if the considered sample is rich enough, which is not yet the case for current three-dimensional observational galaxy catalogs.

Following BSD and BH, our practical implementation $\widetilde{D}_h$ of the asymptotic domain $D_h$ (see eq. [41]) is

$$\widetilde{D}_h = \left\{ N_{\rm c} > 1.8, \quad \ell/\ell_0 < 0.4, \quad N \geq 1, \quad N > 4 N_{\rm v}, \quad N P_N \ell^{-3} > 8 \right\}. \tag{55}$$

The supplementary last condition comes down to removing the very large $N$ part of the CPDF, where this latter is dominated by finite volume effects (c.f. BSD and BH). To enlarge the available dynamic range, we will divide when possible the measured function $\hat{h}$ by the factor $1 + A_{\rm dis}$ computed by BSD, to correct for discreteness effects. This explains why we take $N \geq 1$ instead of $N \gg 1$. Of course, we can still expect significant differences between function $\hat{h}$ and the the searched function $h$, due to the factor $1 + A_{\rm fin}(N, \ell)$.

This section is organized as follows: in § 4.3.1, we measure function $\hat{h}$ in the sets $E_{\rm c}$ and $F$. We want to check if these two reference samples give the expected results, knowing that $F$ is obeying the scaling relation and that $E_{\rm c}$ does not verify it. In § 4.3.2, we measure function $\hat{h}$ in the CDM sample $E$, and its various subsamples randomly extracted. The idea is to dilute $E$ to reach the same number density $n$ as in current observational galaxy samples. This section is concluded by

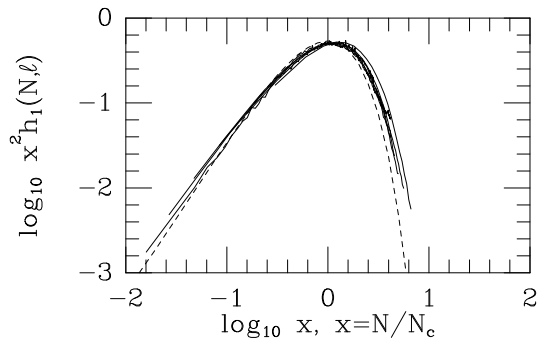

Figure 10: Function $x^2 \hat{h}(N,\ell) \equiv N^2 P_N(\ell)/(n\ell^3)$ as a function of $x \equiv N/N_c$ as measured in the fractal $F$ in logarithmic coordinates. Only the values $(N,\ell) \epsilon \widetilde{D}_h$ (where $\widetilde{D}_h$ is defined by equation [55]) have been selected. The dashed curve is the theoretical expectation $h(x) = 4\exp(-2x)$.

§ 4.3.3, where we give constraints on $n$ to detect a function $h$ (when it exists) and to fairly measure it.

### 4.3.1 The reference samples

The samples $E_c$ and $F$ involve a large number of points $N_{\text{tot}} = 32\,768$. In these sets, we have $\ell_c/\ell_0 \simeq 0.016$, so the effective dynamic range $\widetilde{D}_h$ should be large.

Figure 10 displays the quantity $x^2\hat{h}(N,\ell)$ ($n = 32,768$) measured on $F$ as a function of $x = N/N_c$ for $(N,\ell)\epsilon D_h$ (we take here $\omega = 1$, as suggested by the theoretical calculation of Appendix A and the measurement of function $\sigma$ in § 3.3, so $N_v \equiv 0$). Note that if we took all the available values of $(N,\ell)$, the result would be very similar, which explains why we do not show the corresponding panel. The function $h$ is unambiguously detected: all the curves have the same shape and superpose very well. The dashed line is the theoretical expectation, which is in a good agreement with the measurement.

Left panel of Fig. 11 shows the quantity $x^2 \hat{h}(N,\ell)$ ($n = 32\,768$) measured in $E_c$ as a function of $x = N/N_c$. As expected, the curves are not significantly more gathered than in Fig. 3, which is a first indication against the existence of a function $h$. Right panel of Fig. 11 is the same as left panel, but only the values of $(N,\ell)$ that belong to $\widetilde{D}_h$ have been taken into account. Here, we have not corrected for discreteness, since the value of $\omega = 0$ we get from function $\hat{\sigma}$ (§ 3.3) would call for a special treatment that we felt not necessary to write out. We have however taken $N \geq 2$ instead of $N \geq 1$ to remove the regime completely dominated by discreteness. The scattering of the curves is much smaller in right panel of Fig. 11 than in left panel, which is not surprising since the available dynamic range has been reduced. This could give the misleading illusion that a function $h$ exist, but this scattering is about an order of magnitude, much larger than the expected maximal scattering $S$ that can be infered from equation (53). We indeed find $\log_{10} S = \max[\log_{10}(1 + A_{\text{dis}})] \sim 0.2$ in the scale range given by equation (55). We do not take into account here possible finite volume effects (that are in some way an intrinsic feature of this sample that contains only one cluster). In other words we assume $1 + A_{\text{fin}} \simeq 1$. The computation of $1 + A_{\text{fin}}$ indeed needs to guess a possible function $h$ (eq. [54]), which is quite difficult for this particular case, as can been seen on Fig. 11.

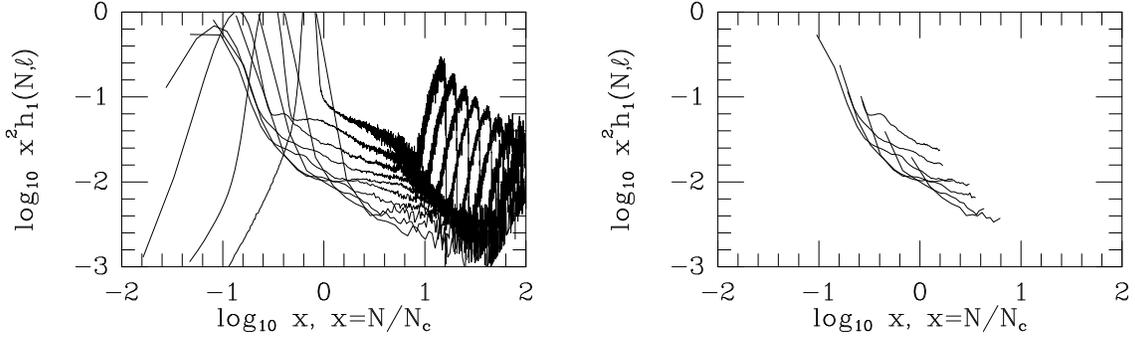

Figure 11: Function $x^2 \hat{h}(N, \ell) \equiv N^2 P_N(\ell)/(n\ell^3)$ as a function of $x \equiv N/N_c$ as measured in the set $E_c$ in logarithmic coordinates. In left panel, all the available values of $(N, \ell)$ have been taken into account. In right panel, we have selected the values $(N, \ell) \epsilon \widetilde{D}_h$, where $\widetilde{D}_h$ is defined by equation (55).

### 4.3.2 A realistic case: dilution effects

Figure 12 displays the quantity $x^2 \hat{h}(N, n, \ell)$ as a function of $x = N/N_c$ for our CDM sample $E$ and various diluted subsamples extracted at random, involving from 32 768 to 512 objects. In left panels, all available values of $(N, \ell)$ are represented. In right panels of Fig. 12, only the values $(N, \ell) \epsilon \widetilde{D}_h$ have been taken into account, and function $\hat{h}$ has been divided by the factor $1 + A_{\text{dis}}$ computed by BSD to correct for discreteness effects. The dashed curve on right panels is the phenomenological fit proposed by BSD (eq. [47]).

In $E$, we have $\ell_c/\ell_0 \simeq 0.016$, as in $E_c$ and $F$. To some extent, one can therefore compare $E$ to $E_c$ and $F$. The curves in right panels of Fig. 12 have all the same regular shape and superpose quite well. Their scattering is of the same order as in Fig. 10 and hence smaller than in right panel of Fig. 11: the existence of a function $h$ is unquestionable. However, when $E$ is randomly diluted, the dynamic domain $\widetilde{D}_h$ is reduced, as well the range of sampled values of $x = N/N_c$. In the subsamples $E_{1024}^{\text{ran}}$ and $E_{32768}^{\text{ran}}$, $\widetilde{D}_h$ is empty. Thus, when the average number density $n$ decreases, the function $\hat{h}$ tends to behave less and less as a function $h$, mainly because of the $1 + A_{\text{dis}}$ factor, that becomes larger and larger. The consequence is that the measurement lies higher and higher above the dashed curve in Fig. 12 (even when one divides the measured function $\hat{h}$ by the factor $1 + A_{\text{dis}}$ computed by BSD; this is because the latter is only approximate; see also Appendix C).

### 4.3.3 Comments

The above examples prove to a large extent that, when they are carefully tested, the predictions of BS on function $\hat{h}$ can be discrimated against. But this is true only if $\ell_c \ll \ell_0$, or equivalently if $N_c(\ell_0) \gg 1$. The measurements on $E$ and its subsamples show that, to detect a function $h$, we must practically have $N_c(\ell_0) \gtrsim 30 - 50$, as it was the case to determine function $\sigma(N_c)$, a condition hardly fulfilled by current three-dimensional galaxy catalogs. To fairly determine function $h$, much more information is needed and $N_c(\ell_0) \gtrsim 80 - 120$ is required. In terms of galaxy number density, this corresponds to

$$n \gtrsim 0.04 - 0.06 h^3 \text{Mpc}^{-3}. \tag{56}$$

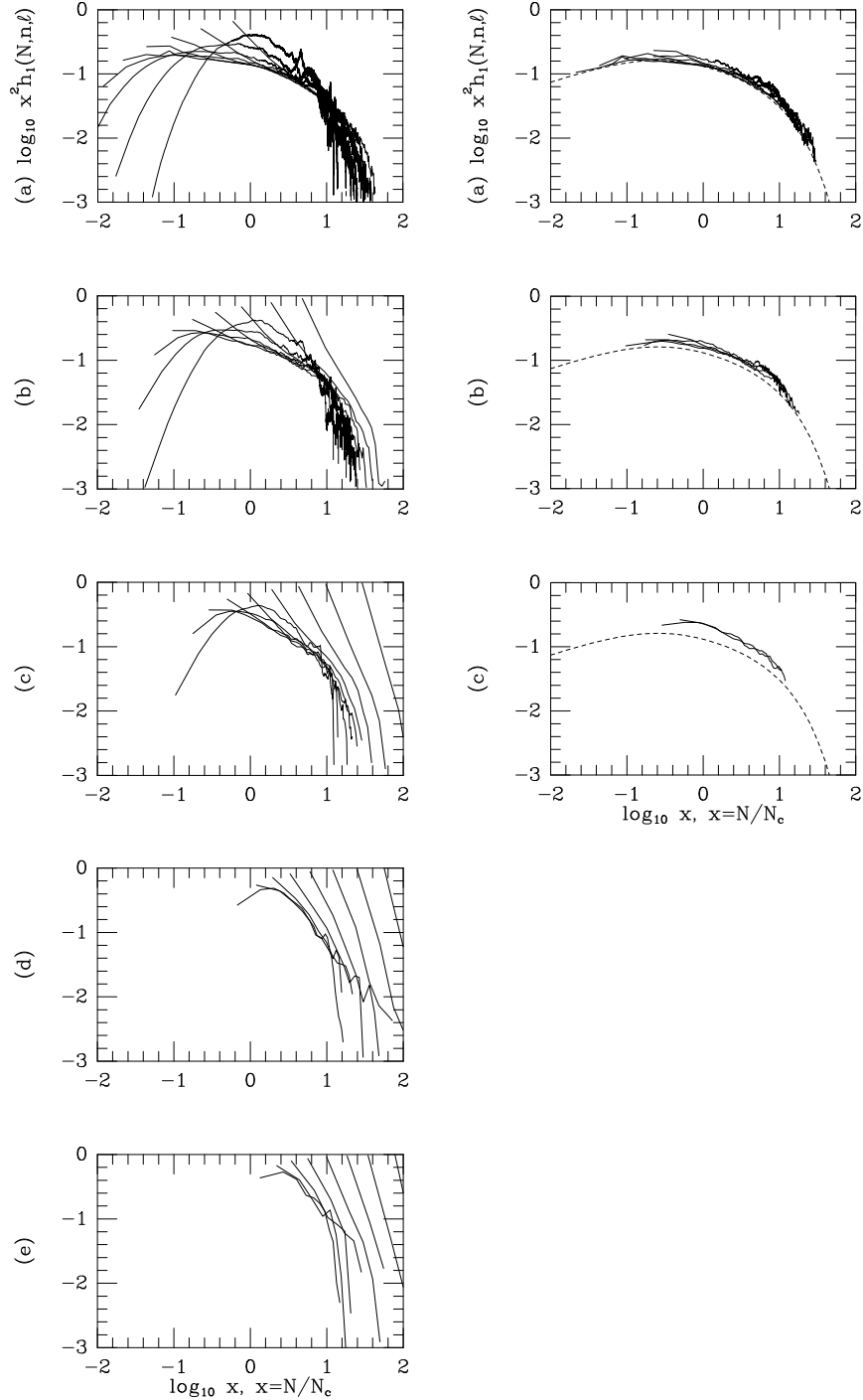

Figure 12: Logarithm of $x^2 \hat{h}(N, n, \ell)$ as a function of $\log_{10} N$ as measured on $E$ (a), $E^{\text{ran}}_{32768}$ (b), $E^{\text{ran}}_{6634}$ (c), $E^{\text{ran}}_{1024}$ (d) and $E^{\text{ran}}_{512}$ (e). All the available values of $(N, \ell)$ are used in left panels. In right panels, only the values of $(N, \ell) \epsilon \widetilde{D}_h$ have been taken into account. For $E^{\text{ran}}_{1024}$ and $E^{\text{ran}}_{512}$, $\widetilde{D}_h$ is empty. We have corrected for discreteness effects at small $N$. The dashed curve is the analytical fit of BSD (eq. [47]). A shift increasing with decreasing number density can be observed between the measurement and this dashed curve. It is essentially due to the fact that $\ell_c$ is approaching $\ell_0$ while the sample is diluted, which reduces the domain $\widetilde{D}_h$ where the behavior of function $\hat{h}(N, n, \ell)$ as a function $h(N/N_c)$ is expected.

Moreover, the size of the sample has to be large in comparison with the correlation length, otherwise finite volume effects will make the exponential tail presented by function $h$ at large $x$ undetectable. Constraint (56) is quite stronger than equation (40), which definitely argues against sparse sampling strategies. Of course, the specific case of our CDM sample cannot be generalized to arbitrary set of points. The work of Maurogordato et al. (1992) on the SSRS catalogue provides rather good indications of the scaling of the CPDF measured in this sample as a function $h$, although their volume limited samples had $n$ at most of order only a few $10^{-3}h^{-3}$ Mpc$^3$. But because $\ell_c$ is close to $\ell_0$, they use a domain $\widetilde{D}'_h$ larger than which is given by equation (55). As it would be the case if we measured the function $\hat{h}$ in $E_{512}^{\rm ran}$ and $E_{1024}^{\rm ran}$ [see panels (e) and (f) of Fig. 12] in $\widetilde{D}'_h$, their function $\hat{h}$ is quite noisy, since the scaling as a function $h$ is expected to be less good in this case. Furthermore, because of the small size of their volume limited sample, their measurement is quite contaminated by finite volume effects.

## 5 Summary and conclusions

We have measured the count probability in three samples, i.e., a power-law spherical cluster $E_c$ plunged in a poissonian distribution involving 32 768 objects, a fractal $F$ generated by a Rayleigh-Lévy random walk involving 32 768 objects, and a CDM sample $E$ involving 262 144 matter particles.

We have evaluated errors due to various spurious effects, such as finite volume effects, discreteness effects and "grid" effects on the void probability (VPDF) and on the count probability (CPDF). Our study has been made in the framework of the scaling model. We indeed wanted to test the practicability and the viability of the formalism of BS. The main results are the following:

### 5.1 About the VPDF and the function $\hat{\sigma}$

1. The VPDF exhibits little sensitivity to finite sample effects, except at the larger scales where it rapidly deteriorates. The trustable scales are smaller than the scale $\ell_{\rm cut}$ defined by

$$P_0(n, \ell_{\rm cut})V_{\rm sample}\ell^{-3} = 1, \qquad (57)$$

where $V_{\rm sample}$ is the sample volume. Above this scale, the measurement of the VPDF is not statistically significant. For $\ell \ll \ell_{\rm cut}$ the error on function $\hat{\sigma} \equiv n^{-1}\ell^{-3}\ln P_0$ should be roughly smaller than the uncertainty on the average number density

$$\frac{\Delta n}{n} = \left[(nV_{\rm sample})^{-1} + \overline{\xi}_2(L_{\rm sample})\right]^{1/2}, \qquad (58)$$

where $\overline{\xi}_2(L_{\rm sample})$ is the average of function $\xi(\mathbf{r}_1 - \mathbf{r}_2)$ over the sampled volume. This error is expected to be very small.

2. However, when one wants to test the scaling model, the quantity $\hat{\sigma}$ is studied as a function of $N_c = n\ell^3\overline{\xi}_2$. This scaling number $N_c$ is rather sensitive to finite volume effects. It is indeed likely to be systematically underestimated by a direct measurement at large scales (CBS). The finite volume error on this number can be approximated by

$$\left(\frac{\Delta N_c}{N_c}\right)^2 \simeq [1 + S_4/4]\overline{\xi}_2(L_{\rm sample}). \qquad (59)$$

This evaluation assumes that the hierarchical model applies (see Appendix B.3). It does not give the amount of the systematic underestimation quoted above but rather estimates the uncertainty on the overall normalization of $N_c$.

3. In numerical simulations, a supplementary spurious effect is expected if the initial distribution of particles was a slightly perturbed regular pattern. This way of setting initial conditions, good from a numerical point of view, because it minimizes white noise at small scales, is bad from a statistical point of view, since it brings some unexpected correlations, that we call grid effects. The consequence is that the VPDF is underestimated, particularly at large scales. Grid effects should be negligible only if

$$P_0(n, \ell) \gtrsim e^{-1}, \qquad (60)$$

which is a rather severe restriction.

4. Once all spurious effects are known and have been isolated, the scaling model can be trustfully tested on the function $\hat{\sigma}$. The procedure consists in measuring the quantity $\hat{\sigma}(n, \ell)$ as a function of $N_c$ in the sample and in various subsamples randomly extracted. This can also be done analytically, using equation (27). Then, if the scaling relation applies in a given scale range, that can be determined by carefully measuring the low-order correlations (CBS, § 4.2.1), an unique function $\sigma(N_c)$ can be determined.

5. To have significant information on the asymptotic behavior of function $\hat{\sigma}$, larges values of $N_c$ are needed. We find with practical measurements that $N_c$ should at least verify

$$N_c(\ell_0) \gtrsim 30 - 40, \qquad (61)$$

where $\ell_0$ is the correlation length, for which $\overline{\xi}_2(\ell_0) = 1$. In terms of galaxy number density, this gives $n \gtrsim 0.02 - 0.03 h^3 \text{ Mpc}^3$. This number is hardly reached by current volume limited three-dimensional galaxy catalogs. This strongly argues against sparse sampling strategies.

## 5.2 About the CPDF and function $\hat{h}$

If the scaling relation applies, the function $\hat{h}(N, n, \ell) \equiv N_c^2 n^{-1} \ell^{-3} P_N(n, \ell)$ should scale as an universal function $h(N/N_c)$, that is characteristic of the underlying continuous density field and has some specific properties computed by BS. However, this behavior is only asymptotic, for a large number of sampling points, and the function $\hat{h}(N, n, \ell)$ is never exactly equal to function $h(x)$, which makes the detection and the measurement of function $h$ somewhat delicate.

1. A small $N$, discreteness effects dominate so $N \gg 1$ is required. Moreover, underdense regions, that have a different scaling behavior contaminate the measurement, so $N \gg N_v$ is required. But such constraints are difficult to follow in realistic samples. Fortunately, BSD found a way to correct for discreteness and the influence of underdense regions, so that one can enlarge the available dynamic range to $N \gtrsim 1$ and $N \gtrsim N_v$. However, their correction needs some improvements (i.e., must be generalized with weaker hypotheses, Schaeffer et al. 1994).

2. Real sample are of finite volume. The consequence is a sharp cutoff at large $N$ on the CPDF. It implies that low order moments of the CPDF are likely to be underestimated by direct measurements. CBS have proposed a method to correct for such defects, which consists in extending to infinity the exponential tail presented by the CPDF at large $N$. One consequence of finite volume effects is also a change in the overall normalization of function $\hat{h}$, which is larger than expected in the vicinity of $N \sim N_\mathrm{c}$. Unfortunately, the corresponding factor $1 + A_\mathrm{fin}$ between functions $\hat{h}$ and $h$ is easy to evaluate only if the scaling relation is obeyed, since one has to know the function $h$ to compute it.

But once these defect have been carefully taken into account, it is seen that the scaling model formalism can be used. However, to trustfully decide if a function $h$ exists or not, condition (61) is required. To measure this function without having to introduce strong corrections, we must require

$$N_\mathrm{c}(\ell_0) \gtrsim 80 - 120. \tag{62}$$

In terms of galaxy density, we thus must roughly have $n \gtrsim 0.04 - 0.06 h^{-3} \mathrm{Mpc}^3$, a constraint that is not fulfilled by current volume limited three dimensional galaxy catalogs.

## 5.3 Perspectives

Our count probability cook book is far from being complete, since we have not taken into account observational effects, such as selection and redshift effects. Moreover, we have only studied the specific case of the scale invariant model of BS. Other models, such as the lognormal distribution (Coles & Jones 1991), the negative binomial model and the thermodynamic model of Saslaw and Hamilton (1984) also provide very good fits of the measured count probability in the observed galaxy distribution (Crane & Saslaw, 1986, 1988). Actually, all these models were seen by Bouchet et al. (1991, 1993) to become indistinguishable in the scaling regime available in the galaxy distribution. As argued by Bouchet et al., this is certainly because the current galaxy catalogs are too sparse and widely dominated their discrete nature (so conditions [61] and [62] are of course not fulfilled), but this issue has to be further clarified.

**Acknowledgments.** SC is supported by DOE and by NASA through grant NAGW-2381 at Fermilab. Part of this work was done while SC was at the Institut d'Astrophysique de Paris (CNRS), supported by Ecole Polytechnique, and at the Lick Observatory (UCSC), invited by L. Hernquist.

# APPENDIX

## A distribution function of our fractal $F$

Here we compute the statistical properties of our fractal $F$. At first, we compute the $Q$-body correlation functions, redoing in a slightly different way a calculation of Peebles (LSS), and show that this set obeys the scaling relation. Then, we try to evaluate functions $\sigma(y)$ and $h(x)$.

### A.1 Computation of the $Q$-body correlations

The mean number density $n$ of a real fractal generated by an infinite Rayleigh-Lévy random walk diverges. Moreover, such a set has no correlation length, i.e., $\xi_2$ is infinite. By imposing that $F$ is a replicated cubical box involving a finite number of objects, we assure the finiteness of $n$ and $\xi_2$.

To test a more realistic situation, we follow LSS and assume here that our set $F$ is generated by a distribution of finite random walks $W_i$, with a random number $N_i$ of steps, started at a random place of the sample, so that the average number density $n_W$ of chains verifies $n = n_W. <N_i>$.

Mandelbrot (1975) and Peebles (LSS, p. 245) give the probability density $f(r)$ of a displacement $\mathbf{r}$ after any number of steps in a given walk $W_i$. With the notations of § 2.3, we have

$$f(r) = C r^{-\gamma}, \quad \gamma = 3 - \epsilon, \quad \epsilon < 2, \tag{63}$$

with

$$C = \frac{(\epsilon+1)(\epsilon-1)}{2\pi^2} \tan\left[-\frac{\pi}{2}\epsilon\right] \ell_p^{-\epsilon}. \tag{64}$$

In the following, we shall use the notation

$$r_{l,m} = |\mathbf{r}_l - \mathbf{r}_m|. \tag{65}$$

We now want to compute the two-body correlation function. Let $\delta P$ be the probability of finding an object in both of the volume elements $\delta V_1$ and $\delta V_2$. The probability that the two objects have been generated by the same walk $W_i$ is $\delta P_1 = [f(r_{1,2}) + f(r_{2,1})] n \delta V_1 \delta V_2$. The first term says that the object 2 have been generated after the object 1 with a probability $f(r_{1,2})\delta V_2$ (and the probability of existence of object 1 is $n\delta V_1$), but the reverse way is possible (term with $r_{2,1}$). The probability that the two objects have been generated by $W_i$ and $W_j$ with $i \neq j$ is $\delta P_2 = n^2 \delta V_1 \delta V_2$, since $W_i$ and $W_j$ are not correlated each other. By definition of the two-body correlation, we have $\delta P \equiv [1 + \xi_2(r_{1,2})] n^2 \delta V_1 \delta V_2 = \delta P_1 + \delta P_2$. So we get

$$n \xi_2(\mathbf{r}_1, \mathbf{r}_2) = 2C r_{1,2}^{-\gamma}. \tag{66}$$

Let us compute the three-body correlation function. Let $\delta P$ be the probability of finding three objects respectively in the three volume elements $\delta V_1$, $\delta V_2$ and $\delta V_3$. We have then three possibilities. The first one is that the three objects have been generated by the same walk $W_i$; the probability of such an event is $\delta P_1 = [f(r_{1,2})f(r_{2,3}) + \text{cyc.} + \text{rev.}] n \delta V_1 \delta V_2 \delta V_3$ (the term "cyc." take into account all the possible paths that can cross the three objects (three terms), and "rev." indicates that the reverse way is possible). The second possibility is that two objects have been generated by a walk $W_i$ and the other object have been generated by $W_j$ with $i \neq j$; the probability of such an event is $\delta P_2 = [f(r_{1,2}) + f(r_{2,3}) + f(r_{3,1}) + \text{rev.}] n^2 \delta V_1 \delta V_2 \delta V_3$. The third possibility is that the three objects have been generated by three different walks; its probability is $\delta P_3 = n^3 \delta V_1 \delta V_2 \delta V_3$. By definition of the three-body correlation function $\xi_3$, we have $\delta P \equiv [1 + \xi_2(r_{1,2}) + \xi_2(r_{2,3}) + \xi_2(r_{3,1}) + \xi_3(r_{1,2}, r_{2,3}, r_{3,1})] n^3 \delta V_1 \delta V_2 \delta V_3 = \delta P_1 + \delta P_2 + \delta P_3$. So we get

$$\xi_3(\mathbf{r}_1, \mathbf{r}_2, \mathbf{r}_3) = [\xi_2(r_{1,2})\xi_2(r_{2,3}) + \text{cyc. (3 terms)}]/2. \tag{67}$$

Let us compute the $Q$-body correlation function. Let $\delta P$ be the probability of finding $N$ objects respectively in the $Q$ volume elements $\delta V_1,...,\delta V_Q$. As we did for the two-body and the three-body correlation functions, we have to find all the possible combinations of sets $W_i$ and paths in $W_i$. Let $\mathcal{P}$ be the set of partitions of the set $\{1,...,Q\}$. An element of $\mathcal{P}$ is written $i = (i_1,...,i_s)$, with $i_k = \{i_{k,1},...,i_{k,\text{card}(i_k)}\}$. With these definitions, we thus have $i_k \cap i_l = \emptyset$ if $k \neq l$, and $i_1 \cup ... \cup i_s = \{1,...,Q\}$. In the following, we use the notation $\text{card}(i_k) = c_k$. Generalizing the above procedure, we find

$$\delta P = \delta V_1 ... \delta V_Q \sum_{(i_1,...,i_s)\epsilon\mathcal{P}} \prod_{k=1}^{s} \sum_{\text{permutations } j[i_k]} n f_{j_1,j_2} \times ... \times f_{j_{c_k-1},j_{c_k}}, \tag{68}$$

with
$$\sum_{\text{permutations } j[i_k], c_k = 1} n f_{j_1, j_2} \times ... \times f_{j_{c_k-1}, j_{c_k}} \equiv n, \tag{69}$$

and $f_{l,m} \equiv f(r_{l,m})$. Now we have

$$\delta P \equiv n^Q \delta V_1 ... \delta V_Q \sum_{(i_1, ..., i_s) \epsilon \mathcal{P}} \prod_{k=1}^{s} \xi_{c_k}(\mathbf{r}_{i_{k,1}}, ..., \mathbf{r}_{i_{k,c_k}}), \tag{70}$$

with $\xi_1 \equiv 1$. It is then easy to recursively compute the $Q$-body correlation function:

$$\xi_Q(\mathbf{r}_1, ..., \mathbf{r}_Q) = [\xi_2(r_{1,2})\xi_2(r_{2,3})...\xi_2(r_{Q-1,Q}) + \text{permutations in } \{1, ..., Q\}, \text{ (i.e., } Q! \text{ terms)}]/2^{Q-1}. \tag{71}$$

This result was already derived (in the framework of the hierarchical model) by Hamilton & Gott (1988). Our set therefore obeys the scaling relation (2).

## A.2 Computation of the count probability

In the following, we assume that the cells are spherical. To compute the function $P_N(\ell)$, we need at first to calculate the averaged $Q$-body correlation function $\overline{\xi}_Q(\ell)$, given by equation (3). We will then be able to compute functions $\sigma(y)$ and $h(x)$.

### A.2.1 Averaged $Q$-body correlations

Equations (66), (71) imply

$$\overline{\xi}_Q(\ell) = S_Q \overline{\xi}_2^{Q-1}, \tag{72}$$

with

$$S_Q = \widetilde{S}_Q \Phi_Q, \tag{73}$$

$$\widetilde{S}_Q = 2^{1-Q} Q!, \tag{74}$$

$$\Phi_Q = \frac{J_Q}{J_2^{Q-1}}, \tag{75}$$

$$J_Q = \left(\frac{4\pi}{3}\right)^{-Q} \int_S d^3 r_1 ... \int_S d^3 r_Q (r_{12}...r_{(Q-1)Q})^{-\gamma}, \tag{76}$$

and $S$ is the sphere of radius unity. The value of $\Phi_Q$ should be close to unity, as argued by BS and Bernardeau & Schaeffer (1992, hereafter BeS), because the correlation function structure (71) obeys the hierarchical model considered by these authors. In spherical coordinates, equation (76) can be written ($\gamma < 2$ and $Q \geq 3$)

$$J_Q = \frac{2}{[(2-\gamma)(3-\gamma)]^{Q-1}} \left(\frac{3}{2}\right)^Q \int_0^1 dr_2 ... dr_{Q-1} I_1(r_2) I_2(r_2, r_3) ... I_2(r_3, r_{Q-1}) I_1(r_{Q-1}), \tag{77}$$

with

$$I_1(r) = (1+r)^{3-\gamma} - (1-r)^{3-\gamma} - \frac{1}{4-\gamma}\left[(1+r)^{4-\gamma} - (1-r)^{4-\gamma}\right], \tag{78}$$

$$I_2(r_1, r_2) = (3-\gamma)\left[(r_1+r_2)^{2-\gamma} - |r_1 - r_2|^{2-\gamma}\right]. \tag{79}$$

$J_2$ can be analytically estimated (LSS) and is equal to $72/[(3-\gamma)(4-\gamma)(6-\gamma)2^\gamma]$. $J_3$ can be reduced to an uni-dimensional integral, $J_4$ and $J_5$ to bi-dimensional integrals, $J_6$ and $J_7$ to tri-dimensional integrals, and so on. We numerically estimated $J_Q$ for $Q \leq 6$. We take $\gamma = 1.8$, as chosen to construct our fractal $F$. The values are listed in Table 1. At low-order, we see that $\Phi_Q$ is close to unity, but $\Phi_Q - 1$ increases with $Q$.

### A.2.2  Evaluation of function $\sigma(y)$ and function $h(x)$

Using equations (4), (10) and (5), we write $\sigma(y)$ as the series expansion

$$\sigma(y) = \sum_{N=1}^{\infty} (-1)^{N-1} \frac{S_N}{N!} y^{N-1} \tag{80}$$

with $S_1 \equiv S_2 \equiv 1$. For our fractal, this expression reads

$$\sigma(y) = \sigma_0(y) + \delta\sigma(y), \tag{81}$$

with

$$\sigma_0(y) = \left(1 + \frac{y}{2}\right)^{-1}, \tag{82}$$

$$\delta\sigma(y) = \sum_{N=1}^{\infty} (-2)^{1-N} (\Phi_N - 1) y^{N-1}, \tag{83}$$

and $\Phi_1 \equiv 1$.

The problem is now to evaluate $\delta\sigma(y) = 0.028(y/2)^2 - 0.064(y/2)^4 + \mathcal{O}(y^5)$. In the regime $y \ll 1$, we have, as expected, $|\delta\sigma(y)| \ll \sigma_0(y)$, which is much less obvious for $y \gtrsim 1$.

However, the quantity $\Phi_Q$ seems to be well approximated by the following phenomenological fit

$$\Phi_Q \sim \Phi_3 \delta^{Q-3}, \quad 3 \leq Q \leq 6, \tag{84}$$

with $\Phi_3 \sim 1.028$ and $\delta \sim 1.036$ (see Table 1). Let us suppose that, in a first approximation, equation (84) is valid for any $Q$. With the above assumptions, we have

$$\sigma(y) \simeq \frac{\Phi_3/\delta^2}{1 + \delta y/2} + (1 - \frac{\Phi_3}{\delta^2}) - (1 - \frac{\Phi_3}{\delta})\frac{y}{2}. \tag{85}$$

With our numerical values we find that

$$\sigma(y) \simeq \frac{0.958}{1 + y/1.93} + 0.0422 - 0.0039y. \tag{86}$$

This expression is practically valid for $y \lesssim 10$. In this regime, we have $|\delta\sigma| \ll \sigma_0$.

But equation (86) was derived in a somewhat ad-hoc way and it does not sample well the large values of $y$. A better way of evaluating the difference between $\sigma$ and $\sigma_0$ consists in using the method explained in appendix A of BeS. This method is valid here since our fractal obeys the hierarchical model used by BeS. Let us rewrite their equations (A13), (A14), (A15), (A16) as

$$\sigma(y) = \frac{1}{v} \int_v d^3r\, s(\mathbf{r}), \tag{87}$$

where $v = \ell^3$ is the volume of a (spherical) cell centered on origin,

$$s(\mathbf{r}) = \zeta[\tau(\mathbf{r})] - \frac{1}{2}\tau(\mathbf{r})\zeta'[\tau(\mathbf{r})], \tag{88}$$

where $\zeta'(\tau) \equiv d\zeta/d\tau$,

$$\tau(\mathbf{r}) = -\frac{y}{v}\int_v d^3r' \frac{\xi_2(\mathbf{r},\mathbf{r}')}{\overline{\xi}_2}\zeta'(\tau(\mathbf{r}')). \tag{89}$$

The function $\zeta(\tau)$ depends on the hierarchy of the $Q$-body correlation functions. Using equations (2), (4) and (5) of BeS, we find, for our fractal,

$$\zeta(\tau) = 1 - \tau + \frac{1}{4}\tau^2. \tag{90}$$

We now want to show that function $\sigma(y)$ is not very different from $\sigma_0(y)$ for any value of $y$. Using (90), we see that function $s$ obeys the implicit equation

$$s(\mathbf{r}) = 1 - \frac{y}{2v}\int_v d^3r' \frac{\xi_2(\mathbf{r},\mathbf{r}')}{\overline{\xi}_2} s(\mathbf{r}'). \tag{91}$$

It is then useful to write $s(\mathbf{r})$ and $\xi_2(\mathbf{r},\mathbf{r}')$ in a perturbative way:

$$s(\mathbf{r}) = s_0 + \delta s(\mathbf{r}), \tag{92}$$

$$\xi_2(\mathbf{r},\mathbf{r}') = \overline{\xi}_2[1 + \delta\xi(\mathbf{r},\mathbf{r}')]. \tag{93}$$

Using equations (87), (91), one of course gets

$$s_0 = \sigma_0 \tag{94}$$

$$\delta s(\mathbf{r}) = -\frac{y}{2v}\int_v d^3r'\delta s(\mathbf{r}') - \frac{y}{2v}\sigma_0\int_v d^3r'\delta\xi(\mathbf{r},\mathbf{r}') - \frac{y}{2v}\int_v d^3r'\delta\xi(\mathbf{r},\mathbf{r}')\delta s(\mathbf{r}'), \tag{95}$$

$$\delta\sigma = \frac{1}{v}\int_v d^3r\,\delta s(\mathbf{r}). \tag{96}$$

Equations (95) and (96) permit to recursively compute $\delta\sigma$:

$$\delta\sigma(y) = \sigma_0 \frac{\Delta_2 \Sigma(y)\frac{y^2}{4}}{1 + \frac{y}{2} + \Delta_2\Sigma(y)\frac{y^3}{8}}, \tag{97}$$

where

$$\Sigma(y) = \sum_{N=0}^{\infty} \frac{\Delta_{N+2}}{\Delta_2}\left(\frac{y}{2}\right)^N, \tag{98}$$

$$\Delta_{N+2} = \frac{(-1)^N}{v^{N+3}}\int_v d^3r_1...d^3r_{N+3}\delta\xi_2(r_{1,2})...\delta\xi_2(r_{N+2,N+3}) = \sum_{k=0}^{N+2}(-1)^{N+k}\frac{(N+2)!}{(N+2-k)!k!}\Phi_{N+3-k}. \tag{99}$$

Numerically, we find $\Sigma(y) = 1 + 0.71(y/2) + 0.54(y/2)^2 + 0.36(y/2)^3 + \mathcal{O}(y^4)$ and $\Delta_2 = \Phi_3 - 1 = 0.028$. If we neglect the very improbable case in which we would have $\Sigma(y) \simeq -(8/\Delta_2)(1 + y/2)y^{-3}$ for $y \gtrsim 1$, we see through equation (97) that $\sigma$ should be always very close to $\sigma_0$, even when $y$ is arbitrarily large, which is confirmed by Fig. 6.

Table 1: Values of $\Phi_Q$, $Q \leq 6$ compared with $\Phi_3 \delta^{Q-2}$ for $\gamma = 1.8$.

| $Q$ | $\Phi_Q$ | $\Phi_3 1.036^{Q-3}$ |
|---|---|---|
| 3 | 1.028 | 1.028 |
| 4 | 1.064 | 1.065 |
| 5 | 1.103 | 1.103 |
| 6 | 1.145 | 1.143 |

In this particular example, the approximation $\Phi_Q \simeq 1$ is thus very good. It is now easy to compute function $h(x)$ from equations (11), (82):

$$h(x) \simeq 4\exp(-2x) \qquad (100)$$

The above reasoning can also be applied to cubical cells (the quantities $\Phi_Q$ are simply slightly different), for which equations (82) and (100) also apply.

## B  Misleading effects on function $\hat{\sigma}$

### B.1  Grid effects and void probability in numerical simulations

From a pure statistical point of view, the discrete realization of a smooth continuous density field should be locally poissonian. On the other hand, it is useful to start a simulation from a slightly perturbed regular pattern of particles in order to reduce the poissonian noise at small scales, which is contradictory with the statistical vision. For instance, our CDM sample comes from a $64 \times 64 \times 64$ initial grid of particles. Now underdense regions, where shell-crossing is not likely to take place, may have conserved the information linked to the initial grid, as can be seen on left panel of Fig. 1. Let us see if this modifies the behavior of the VPDF.

To do that, let us dilute $E$ in two different ways (a) and (b), the first one preserving the possible information associated to a grid, the second one destroying this information. The method to obtain (a) consists in diluting $E$ so that the remaining particles form, at the beginning of the simulation, a (slightly perturbed) regular pattern, i.e., a $m \times m \times m$ grid. We call such a dilution a *preventive* one, since it preserves *all* the properties of the particle distribution (but not all the information). To obtain (b), we simply have to randomly dilute $E$. To significantly destroy the possible information linked to the initial grid, the dilution factor $n/n_i = 64^3/m^3$ must be important, but not too large, otherwise the subsamples (a) and (b) will become indistinguishable, because both dominated by discreteness (i.e., $\hat{\sigma} \sim 1$). We took $n/n_i = 8$ (so $m = 32$). We thus extracted a subsample $E^{\text{grid}}_{32768}$ obeying requirement (a) and three subsamples $E^{\text{ran},j}_{32768}$, $j = 1, 2, 3$, obeying requirement (b).

Figure 13 gives $\hat{\sigma}$ as a function of $N_c$ as measured on $E^{\text{ran},j}_{32768}$ (solid curves), $E$ (dashed curve) and $E^{\text{grid}}_{32768}$ (filled symbols). Clearly, the functions $\hat{\sigma}$ measured on $E^{\text{grid}}_{32768}$ and on $E^{\text{ran},j}_{32768}$ differ at large $N_c$ (so at large scale).

One can evaluate the scale for which the random dilution differs from the preventive one, by comparing the VPDF of a poissonian set with the VPDF of a grid of same average density. In the

first case we get

$$P_0^{\text{ran}}(n,\ell) = \exp(-n\ell^3). \tag{101}$$

In the second case we have

$$P_0^{\text{grid}}(n,\ell) = \max(1 - n\ell^3, 0). \tag{102}$$

Phenomenologically, we see that the statistics of a grid becomes quite different from the statistics of a pure random set when $n\ell^3 > 1$, that is when $P_0^{\text{ran}}(\ell) < 1/e$. This simple comparison also explains why the function $\hat{\sigma}$ measured in $E_{32768}^{\text{grid}}$ is smaller than the one measured in $E_{32768}^{\text{ran},j}$, since a random dilution leads to a greater void probability than a preventive dilution.

Let us look now at the more complicated case of our CDM sample, which is a discrete realization of a continuous underlying density field that can be described by a density contrast $\delta(\mathbf{r})$. For a "large enough" scale, the void probability will be determined by underdense regions, in which $\delta(\mathbf{r})$ is expected to present smooth variations and to have conserved the information on the initial grid. Equations (101) and (102) then become

$$P_0^{\text{ran}}(n,\ell) \simeq \frac{1}{V} \int_V \exp(-[1+\delta(\mathbf{r})]n\ell^3) d^3r, \tag{103}$$

$$P_0^{\text{grid}}(n,\ell) \simeq \frac{1}{V} \int_V \max(1-[1+\delta(\mathbf{r})]n\ell^3, 0) d^3r. \tag{104}$$

$P_0^{\text{grid}}$ vanishes when $[1+\delta(\mathbf{r})]n\ell^3 \geq 1$ for any $\mathbf{r}$. Then, $P_0^{\text{ran}}(\ell) \leq 1/e$. Therefore, it is reasonable to think that when the measured void probability is less than $1/e$, or equivalently, when

$$N_v > 1, \tag{105}$$

grid effects can significantly affect the measurements. In Fig. 13, the points verifying $N_v > 1$ are circled (expect for $E$).

## B.2 Finite sample effects and void probability

### B.2.1 Calculation of the error due to the sample finiteness

We aim here to estimate the error or the VPDF associated to finite sample effects. Let us consider a set $S_{\text{sub}}$ of finite volume $V$, involving a finite number $N_{\text{par}}$ of objects. Let us assume that this set is a subsample extracted from an infinite set $S_{\text{inf}}$ of average number density $n$ for which the hierarchical model applies. In this case, the $Q$-body correlation function can be written as a sum of products of $Q-1$ terms in $\xi_2$ (see Bernardeau & Schaeffer 1992, hereafter BeS)

$$\xi_Q(\mathbf{r}_1, ..., \mathbf{r}_Q) \propto \sum \prod_{Q-1} \xi_2(\mathbf{r}_i, \mathbf{r}_j). \tag{106}$$

This expression is a particular writing of the scaling relation. We will assume, in the following that the cells are spherical of volume $v = 4\pi\ell^3/3 \ll V$.

The VPDF is practically computed by randomly throwing cells in $S_{\text{sub}}$ and by counting the fraction of empty cells. Let $\widetilde{P}_0$ be a statistical indicator of $P_0$. It can be written

$$\widetilde{P}_0 = \frac{1}{C_{\text{tot}}} \sum_i \delta_D(N_i), \tag{107}$$

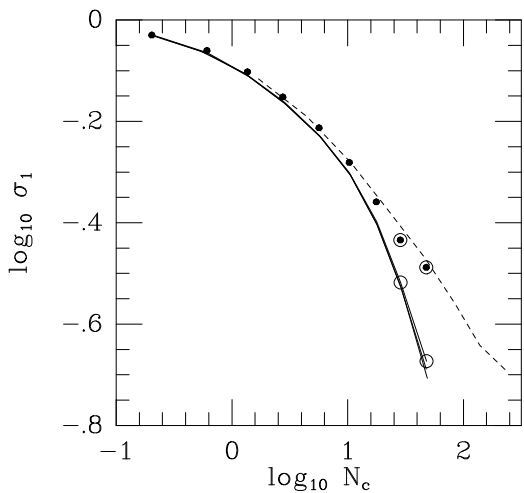

Figure 13: Logarithm of $\hat{\sigma}$ as a function of $\log_{10} N_c$ as measured in $E$ (dashed curve), in three samples $E^{\mathrm{ran},j}_{32768}$, $j = 1, 2, 3$ (solid curves) and in $E^{\mathrm{grid}}_{32768}$ (filled symbols). Except for the dashed curve, the points verifying $P_0 < 1/e$ are circled.

where $N_i$ is the number of objects in the cell $i$, $1 \leq i \leq C_{\mathrm{tot}}$, and $\delta_D(N)$ is the discrete Dirac function that gives 1 if $N = 0$ and 0 otherwise. We assume here that the error related to counts in cells is not significant, or in other words that $C_{\mathrm{tot}}$ is very large. The real VPDF will be the ensemble average of $\widetilde{P}_0$ on an infinite number of realizations $S_{\mathrm{inf}}$:

$$P_0 = <\widetilde{P}_0>_{\mathrm{ens}}. \tag{108}$$

The standard deviation on $P_0$ is written

$$(\Delta P_0)^2 = <\widetilde{P}_0^2>_{\mathrm{ens}} - <\widetilde{P}_0>_{\mathrm{ens}}^2. \tag{109}$$

We have

$$<\widetilde{P}_0^2>_{\mathrm{ens}} = <\frac{1}{C_{\mathrm{tot}}^2} \sum_{i,j} \delta_D(N_i)\delta_D(N_j)>_{\mathrm{ens}}. \tag{110}$$

This quantity is, in the limit $C_{\mathrm{tot}} \gg 1$, nothing but the probability $P_{0,0}(n,\ell,V)$ that two cells thrown at random in $S_{\mathrm{sub}}$ are empty. We can write

$$P_{0,0}(n,\ell,V) = \frac{1}{V^2} \int_V P_{0,0}(n,\ell,r_{1,2}) d^3r_1 d^3r_2, \tag{111}$$

where $P_{0,0}(n,\ell,r)$ is the probability that two cells of volume $v = \ell^3$ separated by a distance $r$ are empty (isotropy is assumed) and $r_{1,2} \equiv |\mathbf{r}_1 - \mathbf{r}_2|$.

Using the results in Appendix C of BeS, we can estimate $P_{0,0}(n,\ell,r)$ in the case $r \gg \ell$ (or equivalently $\xi_2(r) \ll \overline{\xi}_2$). With the notations of BeS, we have

$$P_{0,0}(n,\ell,r) = \chi(0,0). \tag{112}$$

From their equations (A21), (C17), (C18), (C19), we can easily find

$$P_{0,0}(n,\ell,r) = \exp\left\{-2nv[\sigma(N_c) + nv\sigma'(N_c)\xi_2(r)]\right\}, \quad r \gg \ell \tag{113}$$

where $\sigma'(N_c) \equiv d\sigma/dN_c$. When $r \lesssim 2\ell$, the estimation of $P_{0,0}(n,\ell,r)$ is rather complicated, since the two cells $C_i$ and $C_j$ are connected, forming a non-spherical cell of volume

$$v'(r) \equiv \frac{4}{3}\pi[\ell'(r)]^3 = 2v - \int_{C_i \cap C_j} d^3r = v\left[1 + \frac{3}{4}\frac{r}{\ell} - \frac{1}{16}\left(\frac{r}{\ell}\right)^3\right]. \tag{114}$$

But let us assume, to simplify the calculation, that the VPDF does not strongly depend on the shape of the cell (which is true in the poissonian case). Then, we can write

$$P_{0,0}(n,\ell,r) \simeq P_0(n,\ell'(r)), \quad r \lesssim 2\ell. \tag{115}$$

The integral (111) then becomes

$$P_{0,0}(n,\ell,V) \simeq \frac{4\pi}{V}\int_0^{2\ell} P_0(n,\ell'(r))r^2 dr + \frac{P_0^2}{V^2}\int_{r_{1,2} \gtrsim 2\ell} \exp\left\{-2(nv)^2\sigma'\xi_2(r_{1,2})\right\} d^3r_1 d^3r_2. \tag{116}$$

The first integral and the second integral of the right hand side of this equation will be denoted by $I_1$ and $I_2$. Let us assume that $\xi_2(r)$ is a power-law of index $-\gamma$ where $3/2 < \gamma < 3$. $I_2$ can be approximated by

$$I_2 \simeq P_0^2\left(1 - \frac{8v}{V} + \eta\bar{\xi}_2(L) + \frac{4\pi}{V}\int_{2\ell}^{+\infty} \sum_{N \geq 2} \frac{[\eta\xi_2(r)]^N}{N!} r^2 dr\right), \tag{117}$$

where $\bar{\xi}_2(L)$ is the averaged correlation over the sample:

$$\bar{\xi}_2(L) \equiv \frac{1}{V^2}\int_V \xi_2(r_{1,2}) d^3r_1 d^3r_2 \tag{118}$$

and

$$\eta \equiv -2(nv)^2\sigma'(N_c). \tag{119}$$

Integrating (117) leads to

$$I_2 \simeq P_0^2\left(1 - \frac{8v}{V} + \eta\bar{\xi}_2(L) + \frac{4\pi(2\ell)^3}{V}\sum_{N \geq 2} \frac{[\eta\xi_2(2\ell)]^N}{(\gamma N - 3)N!}\right). \tag{120}$$

In the case $1 < \gamma \leq 3/2$, the result is similar. The difference is that a supplementary term $P_0^2\eta^2 V^{-2}\int_{r_{1,2} \gtrsim 2\ell} \xi_2^2(r_{1,2}) d^3r_1 d^3r_2$ replaces the $N = 2$ term of the sum in right hand side term of equation (120) that starts at $N = 3$ instead of $N = 2$. In the following, we still assume that $3/2 < \gamma < 3$, although we think that the final result can be reasonably generalized for $1 < \gamma \leq 3/2$.

We now want to find reasonable estimates of $I_1$ and $I_2$. To do that, we consider two extreme cases. The first one (i) corresponds to the regime when $P_0 \sim 1$, or equivalently $nv\sigma \sim 0$. The second one (ii) corresponds to the regime when $P_0 \ll 1$ or equivalently $nv\sigma \gg 1$. We will then consider two subcases (a) and (b) in (i) and (ii), respectively the "poisson limit" $N_c \ll 1$ ($\sigma \sim 1$) and the asymptotic regime (22) expected when $N_c \gg 1$.

(ia) $P_0 \sim 1$, $N_c \ll 1$: in this case, it is easy to compute $I_1$ and $I_2$ at first order in $N_c$ and $nv\sigma$. We find

$$I_1 \simeq \left[P_0 + \frac{nv}{8}\right]\frac{8v}{V}P_0. \tag{121}$$

Since $\sigma \sim 1 - N_c/2$ when $N_c \ll 1$ (see eq. [80]), we have $\sigma' \simeq -1/2$ and

$$I_2 \simeq \left[1 - \frac{8v}{V} + (nv)^2 \overline{\xi}_2(L)\right] P_0^2. \tag{122}$$

(ib) $P_0 \sim 1$, $N_c \gg 1$: the calculation of $I_1$ at first order in $nv\sigma$ leads to

$$I_1 \simeq \left[P_0 + \frac{\alpha nv\sigma}{8}\right] \frac{8v}{V} P_0, \tag{123}$$

where $\alpha$ is a factor depending on the values of $\omega$ and $\gamma$, but it is numerically seen to be close to unity (or of the order of a few). We can also write

$$\eta \xi_2(2\ell) \simeq 2\omega nv\sigma(N_c) \frac{\xi_2(2\ell)}{\overline{\xi}_2(\ell)}, \quad N_c \gg 1. \tag{124}$$

So $\eta \xi_2(2\ell) \ll 1$ and $I_2$ is given at first order in $nv\sigma$ by

$$I_2 \simeq \left[1 - \frac{8v}{V} - 2(nv)^2 \sigma' \overline{\xi}_2(L)\right] P_0^2. \tag{125}$$

(iia) $P_0 \ll 1$, $N_c \ll 1$: The computation of $I_1$ in this regime is easy. One can find

$$I_1 \simeq \left[P_0 + \frac{128/9}{8(nv)^3}\right] \frac{8v}{V} P_0. \tag{126}$$

We have $\eta \xi_2(2\ell) \propto nv N_c$ and $nv\sigma \simeq nv \gg 1$. Thus, the first order dominant term in equation (120) is also compatible with equation (122) when $nv N_c \ll 1$ (in this case, $I_2 \simeq [1 - 8v/V] P_0^2$ since $\overline{\xi}_2(L) \ll \overline{\xi}_2(\ell) \ll 1$). When $nv N_c \gg 1$, we obtain similar conclusions as in item (iib), but the quantity $B$ defined hereafter is negligible compared with $I_1 P_0^{-2} - 8(v/V)$.

(iib) $P_0 \ll 1$, $N_c \gg 1$: $I_1$ is written

$$I_1 \simeq \left[P_0 + \frac{128/9}{8(\beta nv\sigma)^3}\right] \frac{8v}{V} P_0, \tag{127}$$

where $\beta = 1 - (3 - \gamma)\omega/3$ is close to unity. $I_2$ is given by

$$I_2 \sim \left[1 - \frac{8v}{V} + \eta \overline{\xi}_2(L) + B\right] P_0^2, \tag{128}$$

where $B$ is of the order of

$$B \sim \frac{v}{V} \frac{\exp(\eta \xi_2(2\ell))}{\gamma \eta \xi_2(2\ell)}. \tag{129}$$

With equation (124) which is also valid in this regime, we see that $B P_0^2$ should be of the same order as $I_1 - 8(v/V) P_0^2$ (we cannot make more detailed comparisons, since we are using rough approximations).

In the regime (i) or $nv\sigma \sim 0$, we thus have

$$(\Delta P_0)^2 \simeq \alpha \sigma nv \frac{v}{V} P_0 - 2(nv)^2 \sigma' \overline{\xi}_2(L), \quad P_0 \simeq 1. \tag{130}$$

Numerical calculations moreover show that in the regime $nv\sigma(N_c) \lesssim 1$, $I_1$ is well approximated by

$$I_1 \simeq \left[ P_0 + \frac{\alpha}{8}(1 - P_0) \right] \frac{8v}{V} P_0, \quad P_0 \gtrsim 1/e, \tag{131}$$

where $\alpha$ is a factor (not necessarily constant) of the order of unity or at most a few unities (and $\alpha = 1$ exactly in the Poisson case). When $P_0$ is very small, this expression is of course not valid, and a logarithmic correction should be taken into account [i.e., a term $(nv\sigma)^{-3} = -(\ln P_0)^{-3}$, see equations (126), (127)], but at the level of approximation we are, we shall forget it and we shall take $\alpha = 1$.

The expected error on function $P_0$ should therefore roughly be

$$\left( \frac{\Delta P_0}{P_0} \right)^2 \simeq \frac{v}{V} \frac{1 - P_0}{P_0} - 2(nv)^2 \sigma'(N_c) \overline{\xi}_2(L). \tag{132}$$

Of course, this expression is compatible with equation (130).

The first right hand side term of equation (132) should be valid even if the hierarchical model does not applies. Indeed, it simply says that the number of independent cells in the sample is $V/v$. The second right term is however more questionable. An alternative calculation of this term was made by Colombi (1993), who did not assume anything about the scaling behavior of the underlying distribution, but used a scale reasoning, the spherical top hat approximation and the fact that the distribution was initially gaussian. We simply quote here the result

$$\left( \frac{\Delta P_0}{P_0} \right)^2 \simeq (nv)^2 \hat{\sigma}^{68/21} \overline{\xi}_2(L), \tag{133}$$

which is quite similar (but not equal) to the second right hand sided term of equation (132).

### B.2.2   Rare voids, the largest void

In the approximation (132), we see that the error becomes always larger than unity above the scale $\ell_{\text{cut}}$ defined by

$$P_0(n, \ell_{\text{cut}}) V / v(\ell_{\text{cut}}) \sim 1. \tag{134}$$

This is to be compared with the certainly more accurate estimate (coming from eq. [127])

$$P_0(n, \ell_{\text{cut}}) V / v(\ell_{\text{cut}}) \sim 14.2 [-\ln P_0(n, \ell_{\text{cut}})]^{-3}. \tag{135}$$

Indeed, this equation can be considered as a generalization of a result of Polizer & Preskill (1986), who computed the number $n_{\text{voids}}$ of "rare" void cells in a poissonian sample of volume $V$. Their result is

$$n_{\text{voids}}^{\text{poisson}} = \frac{V}{v} (nv)^3 P_0, \quad P_0 \ll 1. \tag{136}$$

There is typically only one void when $P_0(n,\ell) V / v \sim (nv)^{-3} = [-\ln P_0(n,\ell)]^{-3}$, which is quite similar to equation (135) if we forget the factor 14.2. Equation (136) may be thus generalized for a clustered distribution of points as

$$n_{\text{voids}} \sim \frac{V}{v} [-\ln P_0]^3 P_0, \quad P_0 \ll 1. \tag{137}$$

For our CDM sample, we obtain $\log_{10} \ell_{\rm cut} \simeq -1.3$ from equation (134) and $\log_{10} \ell_{\rm cut} \simeq -1.25$ from equation (135), which are quite similar values and that are moreover in agreement with Fig. 4. Indeed the far right point of this figure corresponds to the measurement of function $\hat{\sigma}$ in $E$ for $\log_{10} \ell = -1.2$. It is abnormally shifted forward, meanwhile the next point on the left, which corresponds to $\log_{10} \ell = -1.4$, seems to be good.

When $\ell \gtrsim \ell_{\rm cut}$, two cases can be considered:

1. If the largest void is slightly too large in comparison with its expected size (that can be obtained by making an average on numerous realizations of volume $V$), the VPDF is artificially overestimated when $\ell > \ell_{\rm cut}$, presents at larger scale an abrupt cut off and vanishes at $\ell_{\rm cutoff} > \ell_{\rm cut}$. In this case, the measured function $\hat{\sigma}$ presents a cutoff between $\ell_{\rm cut}$ and $\ell_{\rm cutoff}$, then increases suddenly toward infinity when $\ell$ tends to $\ell_{\rm cutoff}$. This is illustrated by Fig. 6 (but our logarithmic bin is too large to see the increase on $\hat{\sigma}$).

2. If the largest void is slightly too small in comparison to its expected value, then the VPDF presents only an abrupt cutoff between $\ell_{\rm cut}$ and $\ell_{\rm cutoff}$, which means that the measured $\hat{\sigma}$ increases rapidly to infinity when $\ell \longrightarrow \ell_{\rm cutoff}$. This is the case of $E$, which is contaminated by grid effects that tend to decrease the expected size of the voids.

### B.2.3 Example

Let us see now, through a simple example, if the estimation (132) of the error on the function $P_0$ is realistic. We have extracted from our CDM sample eight adjacent cubical subsamples of size $L_{\rm sample} = L_{\rm box}/2$, in which we have measured the function $\hat{\sigma}(n,\ell) \equiv -\ln(P_0)/n\ell^3$. Figure 14 gives the logarithm of $\hat{\sigma}$ as a function of $\log_{10}(\ell)$ (triangles refer to the main sample, and curves to the subsamples). The far right square indicates that $|\Delta P_0/P_0| > 1$. To estimate $\overline{\xi}_2(L_{\rm sample})$ we have directly measured it in our CDM sample, and we find $\overline{\xi}_2(L_{\rm box}/2) \sim 0.04$. The errorbars in Fig. 14 represent the expected error on function $\hat{\sigma}$ (using eq. [37]). They seem to be of the appropriate size, although slightly too large. But we must be aware of the fact that equation (132) is rigorously valid only if the hierarchical model (106) applies, which is not obviously the case of our CDM sample, that for example exhibit slight deviations from the scaling relation at large scales (see § 3 and § 4.2.1).

### B.3 Finite volume error on $\overline{\xi}_2$

In the following, we neglect discreteness effects. Practically, the two-body correlation is computed in a sample of finite volume $V$, i.e.,

$$\widetilde{\xi}_2(r) = \frac{1}{4\pi V} \int_V \int_{\theta,\phi} \delta(\mathbf{x} + r\mathbf{e}_{\theta,\phi}) \delta(\mathbf{x}) \sin\theta d\theta d\phi d^3 x, \tag{138}$$

where $\mathbf{e}_{\theta,\phi} \equiv (\cos\phi\sin\theta, \sin\phi\sin\theta, \cos\theta)$ and $\delta(\mathbf{r})$ is the density contrast of the underlying distribution. In the following, we shall denote by $< X >$ the ensemble average of quantity $X$. For example, the real two-body correlation of the studied infinite set is

$$\xi_2 = < \widetilde{\xi}_2 >. \tag{139}$$

The average standard deviation is

$$(\Delta \xi_2)^2 = < (\widetilde{\xi}_2 - < \widetilde{\xi}_2 >)^2 >. \tag{140}$$

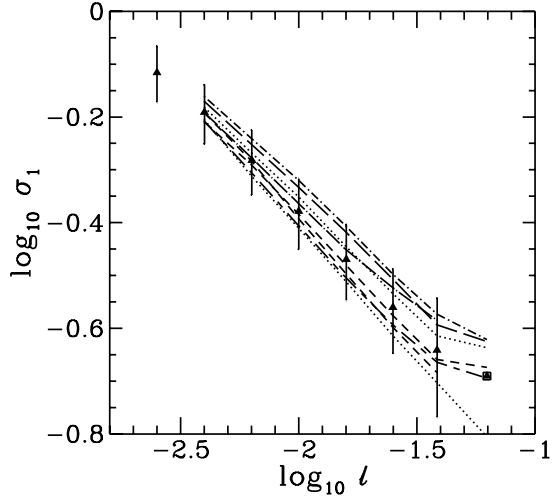

Figure 14: Logarithm of $\hat{\sigma}$ as a function of $N_c$ as measured on our CDM sample $E$ (triangles) and eight adjacent subsamples of size $L_{\text{box}}/2$. The errorbars are given by equations (37), (132). The far right square indicates that $|\Delta P_0/P_0| > 1$.

So we have

$$(\Delta \xi_2)^2 = \frac{1}{(4\pi V)^2} \left\langle \int \delta(\mathbf{x} + \mathbf{r}_1)\delta(\mathbf{x})\delta(\mathbf{y} + \mathbf{r}_2)\delta(\mathbf{y}) \sin\theta_1 d\theta_1 d\phi \sin\theta_2 d\theta_2 d\phi_2 d^3x d^3y \right\rangle - \xi_2^2, \quad (141)$$

with $\mathbf{r}_i = r\mathbf{e}_{\theta_i,\phi_i}$, $i = 1, 2$. If we invert the ensemble averaging and the integral, we get

$$\begin{aligned}(\Delta \xi_2)^2 &= \frac{1}{(4\pi V)^2} \int [\xi_4(\mathbf{x} + \mathbf{r}_1, \mathbf{x}, \mathbf{y} + \mathbf{r}_2, \mathbf{y}) \\ &+ \xi_2(|\mathbf{x} - \mathbf{y} + \mathbf{r}_1 - \mathbf{r}_2|)\xi_2(|\mathbf{x} - \mathbf{y}|) \\ &+ \xi_2(|\mathbf{x} - \mathbf{y} + \mathbf{r}_1|)\xi_2(|\mathbf{x} - \mathbf{y} - \mathbf{r}_2|)] \sin\theta_1 d\theta_1 d\phi_1 \sin\theta_2 d\theta_2 d\phi_2 d^3x d^3y\end{aligned} \quad (142)$$

If the factorized hierarchical model applies [so the system obeys the scaling relation (2)], $\xi_4$ can be written (Fry & Peebles 1978, Sharp et al. 1984)

$$\begin{aligned}\xi_4(1,2,3,4) &= R_a[\xi_2(1,2)\xi_2(2,3)\xi_2(3,4) + \text{cyc. (12 terms)}] \\ &+ R_b[\xi_2(1,2)\xi_2(1,3)\xi_2(1,4) + \text{cyc. (4 terms)}].\end{aligned} \quad (143)$$

Equation (142) is now a sum of various integrals over products of $\xi_2$. The dominant term is, for $r^3/V$ small enough [more precisely for $(r^3/V)(1+\xi_2) \ll \bar{\xi}_2(L)$]

$$(\Delta \xi_2)^2 = 4R_a \xi_2^2 \bar{\xi}_2(L). \quad (144)$$

In the observed galaxy distribution, we have $Q_4 = (12R_a + 4R_b)/16 \sim 2.9 \pm 0.5$ (Fry & Peebles 1978), so the relative error on $\xi_2$ is approximately $\Delta\xi_2/\xi_2 \sim 3.4\sqrt{\bar{\xi}_2(L)}$.

We can also evaluate $Q_4$ as a function of $S_4 = \overline{\xi}_4/\overline{\xi}_2^3$. Using the approximation $S_4 \sim 16Q_4$ (see, e.g., BS), and $R_a \sim R_b$, we find

$$\left(\frac{\Delta\overline{\xi}_2}{\overline{\xi}_2}\right)^2 \simeq \frac{S_4}{4}\overline{\xi}_2(L). \tag{145}$$

# C  Contamination effects on function $h$

In the following, we assume that the scaling relation applies. The domain of validity $D_h$ (eq. [41]) of equation (9) is asymptotic. So practically, $\hat{h}$ never scales exactly as a function $h$. Moreover, some misleading effects can be brought by the finiteness of the sampled volume. So $\hat{h}$ can be given by

$$\hat{h}(N, n, \ell) = U(N, \ell, n).h(x), \tag{146}$$

where $U(N, \ell, n)$ is a factor which tends to unity when $(N, \ell)\epsilon D_h$ and when the size of the sample tends to infinity. It is then useful to separate the different contamination effects that may intervene. So we write

$$U(N, \ell, n) = [1 + A_{\text{dis}}(N, n, \ell)].[1 + A_{\text{fin}}(N, \ell)]. \tag{147}$$

The term $A_{\text{dis}}(N, n, \ell)$ is related to discreteness effects and the contamination by underdense regions The term $A_{\text{fin}}(N, \ell)$ is linked to finite volume effects. These two effects are mixed up, so it is rigorously impossible to treat them separately. Our aim here is simply to roughly estimate them in order to understand the main features of the deviations from the theory brought by these defects. For a detailed evaluation of $A_{\text{dis}}(N, n, \ell)$ as a function of $N$, we refer to the appendix of BSD. Note that, to estimate $A_{\text{dis}}(N, n, \ell)$, BSD assume that the function $\sigma(N_c)$ has reached its asymptotic power-law behavior at large $N_c$. In § C.1, without making such hypothesis, we roughly estimate some average of the number $A_{\text{dis}}$ by using moments of the CPDF. In § C.2, we use a similar reasoning to estimate the change on the overall normalization brought by finite volume effects.

## C.1  Contamination by the finiteness of the available scaling range

Let us assume that our sample is of very large volume, so that finite volume effects are negligible. The calculation of the moment of order two $\nu_2 = \sum N^2 P_N$ of the function $P_N(n, \ell)$ provides

$$\nu_2 = n\ell^3 \sum_{N=1}^{\infty} (N/N_c)^2 \hat{h}(N, n, \ell) = n\ell^3(1 + N_c + n\ell^3), \tag{148}$$

The above sum is dominated by values of $N$ around $N_c$. When $N_c >$ several unities, equality (148) is dominated by values of $N$ larger enough so that the above sum can be replaced by an integral. We can thus write, if $N_c >$ several unities,

$$\nu_2 = \overline{N}N_c \int_0^\infty [1 + A_{\text{dis}}(N_c x, n, \ell)]x^2 h(x) dx. \tag{149}$$

With the definition

$$<A_{\text{dis}}(n, \ell)>_Q \equiv \frac{\int_0^\infty A_{\text{dis}}(N_c x, n, \ell) x^Q h(x) dx}{\int_0^\infty x^Q h(x) dx}, \tag{150}$$

we finally get

$$<A_{\text{dis}}(n, \ell)>_2 = \frac{1}{N_c} + \frac{1}{\overline{\xi}_2}, \tag{151}$$

since function $h(x)$ has to obey the normalizations (46).

Equivalently, we have at first order in $1/N_c$ and $1/\overline{\xi}_2$:

$$\begin{aligned} <A_{\rm dis}(n,\ell)>_3 &\sim \frac{3}{S_3}\left[\frac{1}{N_c} + \frac{1}{\overline{\xi}_2}\right], \\ <A_{\rm dis}(n,\ell)>_4 &\sim \frac{4S_3}{S_4}\left[\frac{1.5}{N_c} + \frac{1+0.75/S_3}{\overline{\xi}_2}\right], \\ <A_{\rm dis}(n,\ell)>_5 &\sim \frac{5S_4}{S_5}\left[\frac{2}{N_c} + \frac{1+2S_3/S_4}{\overline{\xi}_2}\right]. \end{aligned} \quad (152)$$

The calculation of $<A_{\rm dis}(n,\ell)>_Q$ for our CDM universe provides decreasing values of $Q$, which indicates that $A_{\rm dis}(N_c x, n, \ell)$ should be a decreasing function of $x$. Indeed, the larger $Q$, the larger are the contributing values of $x$ to the moments of order $Q$ of the count probability. The variations of $<A_{\rm dis}(n,\ell)>_Q$ with $Q$ depend on the degree of clustering of the system. For example, our fractal $F$, which is much less correlated than $E$ in the sense that its $S_Q$ increase much less rapidly with $Q$, has $<A_{\rm dis}(n,\ell)>_Q$ increasing with $Q$. The dependence on $N$ of $A_{\rm dis}$ is however complex. It has been thoroughly evaluated by BSD in the regime when $\sigma$ reach its asymptotic power-law behavior. The calculation of BSD also includes higher order terms in $1/N_c$ and $1/\overline{\xi}_2$ than equations (151) and (152).

Nevertheless, at $x \sim 1$, which should be in the vicinity of the maximum of the function $x^2 h(x)$, we should roughly have

$$A_{\rm dis}(N, n, \ell) \sim \frac{1}{N_c} + \frac{1}{\overline{\xi}_2}. \quad (153)$$

Thus, the shift between function $\hat{h}$ and function $h$ around $x \sim 1$ should increase while the sample is diluted (because $N_c$ is proportional to the number density $n$), in qualitative agreement with figure 12. Equation (153) also indicates that if we cannot both have $N_c > 1$ and $\overline{\xi}_2 > 1$ in the considered sample, the study of function $h(x)$ becomes difficult and possibly meaningless.

## C.2 Finite volume effects

We assume here that our sample is of finite volume, but also that the above effects are negligible, i.e. that we have $\ell_c \ll \ell \ll \ell_0$ and $N \gg 1$, $N \gg N_v$. Because of the sampled volume finiteness, the maximum number $N_{\rm max}$ of objects that can be found in a cell is necessarily finite, smaller than $N_{\rm par}$, the total number of objects in the set. So the second order moment of the count probability is (in the continuum limit, i.e., $\ell \gg \ell_c$)

$$\nu_2 = \overline{N} N_c = \sum_{N=1}^{N_{\rm max}} N^2 P_N = \overline{N} N_c \int_0^{N_{\rm max}/N_c} [1 + A_{\rm fin}(xN_c, n, \ell)] x^2 h(x) dx, \quad (154)$$

where $N_{\rm max}$ is the maximum number of objects per cell. So we find

$$\int_0^{N_{\rm max}/N_c} [1 + A_{\rm fin}(xN_c, n, \ell)] x^2 h(x) dx = 1. \quad (155)$$

Now we know that function $h$ obeys normalizations (46), so the quantity

$$1 + <A_{\rm fin}(\ell)>_Q \equiv \int_0^{N_{\rm max}/N_c} [1 + A_{\rm fin}(xN_c, n, \ell)] x^Q h(x) dx \Big/ \int_0^{N_{\rm max}/N_c} x^Q h(x) dx \quad (156)$$

is simply, when $Q = 2$,

$$1+ < A_{\text{fin}}(\ell) >_2 = 1 \bigg/ \int_0^{N_{\max}/N_c} x^2 h(x) dx > 1. \tag{157}$$

Since function $x^2 h(x)$ has its maximum at the vicinity of $x \sim 1$, we can expect that for $x \sim 1$, $A_{\text{fin}}(xN_c, \ell) \sim < A_{\text{fin}}(\ell) >_2$. But this is true only if $N_{\max}$ is large enough in comparison with $N_c$ so that $N_c$ is not affected by finite sample effects when $\ell \leq \ell_0$ (which is indeed the case in our CDM sample as shown by CBS). The same calculation can be made for $Q \geq 3$. However, when $Q \geq 3$, one rather tests finite volume effects on quantities $S_Q$ (see, e.g., CBS) than a difference of normalization between functions $\hat{h}$ and $h$. Indeed, the maximum of $x^Q h(x)$ increases with $Q$, so is likely to be of the order of $N_{\max}/N_c$ or even larger.

With regard to our CDM sample, we find that $< A_{\text{fin}}(\ell) >_2 \leq 0.12$ when $\log_{10}(\ell) \leq -1.6$, and $< A_{\text{fin}}(\ell) >_2 \geq 0.26$ when $\log_{10}(\ell) \geq -1.2$. So the shift between $\hat{h}$ and $h$ brought by the volume finiteness is not negligible, and should even more significant for systems having more power at large scales, such as hot dark matter numerical samples.